\documentclass[a4paper,11pt]{article}
\usepackage{jheppub} 
\usepackage{lineno}
\usepackage{circuitikz}
\usepackage{mathtools}
\usepackage{siunitx}
\usepackage{comment}
\usepackage{xcolor}
\usepackage{subfigure}
\usepackage{url}

\arxivnumber{2411.1834} 

\title{\boldmath First characterisation of the MAGO cavity, a superconducting RF detector for kHz-MHz gravitational waves}

\author[a]{Lars Fischer}
\author[b]{Bianca Giaccone}
\author[b]{Ivan Gonin}
\author[b]{Anna Grassellino}
\author[a]{Wolfgang Hillert}
\author[b]{Timergali Khabiboulline}
\author[a]{Tom Krokotsch}
\author[a,c]{Gudrid Moortgat-Pick}
\author[c]{Andrea Muhs}
\author[b]{Yuriy Orlov}
\author[a]{Michel Paulsen}
\author[c]{Krisztian Peters}
\author[b]{Sam Posen}
\author[b]{Oleg Pronitchev}
\author[a,c]{Marc Wenskat}

\affiliation[a]{Universität Hamburg,\\
Luruper Chaussee 149, 22761 Hamburg, Germany}
 \affiliation[b]{Fermi National Accelerator Laboratory,\\
Kirk and Pine St, Batavia, IL 60510, United States}
\affiliation[c]{Deutsches Elektronen-Synchrotron DESY,\\
 Notkestraße 85, 22607 Hamburg, Germany}

\emailAdd{krisztian.peters@desy.de, marc.wenskat@desy.de, giaccone@fnal.gov}

\abstract{Heterodyne detection using microwave cavities is a promising method for detecting high-frequency gravitational waves or ultralight axion dark matter. In this work, we report on studies conducted on a spherical 2-cell cavity developed by the MAGO collaboration for high-frequency gravitational waves detection. Although fabricated around 20 years ago, the cavity had not been used since. Due to deviations from the nominal geometry, we conducted a mechanical survey and performed room-temperature plastic tuning. Measurements and simulations of the mechanical resonances and electromagnetic properties were carried out, as these are critical for estimating the cavity’s gravitational wave coupling potential. Based on these results, we plan further studies in a cryogenic environment. The cavity characterisation does not only provide valuable experience for a planned physics run but also informs the future development of improved cavity designs.}

\begin{document}
\preprint{FERMILAB-PUB-24-0819-SQMS-TD, DESY-24-181}
\maketitle
\flushbottom

\section{Introduction}
\label{sec:intro}

In the search for gravitational waves (GWs) the central focus has been on the Hz to kHz frequency range, which is where the strongest signals from known astrophysical objects were expected. This is the frequency band where the LIGO/Virgo interferometers discovered GWs in 2015 which were produced in the merging of two massive black holes \cite{LIGOScientific:2016aoc}. 
Interferometers were not the only technology developed in the search for GWs. 
Another class of instruments, known as modern Weber bars~\cite{Weber:1960zz}, relied on resonant mechanical bars. Instruments such as ALLEGRO~\cite{Mauceli:1995rm}, AURIGA~\cite{Cerdonio:1997hz}, EXPLORER~\cite{Astone:1993ur}, NAUTILUS~\cite{Astone:1997gi}, and NIOBE~\cite{Blair:1995wx} were the leading detection tools before the advent of interferometers. 
Electromagnetic (EM) cavities can also be employed in the search for GWs, where the mechanical structure of the cavity itself plays the role of the resonant bar. In this setup, the electromagnetic eigenmodes of the cavity serve as mechanical-to-EM transducers, analogous to Weber bars, where the transducer is an LC circuit.

In this detection concept, an electromagnetic resonator is configured with two nearly degenerate modes, where RF power is injected into only one mode. An incoming GW can transfer power from the loaded mode (0) to the quiet mode ($\pi$) which is maximised when the resonant condition $|\omega_\pi - \omega_0| = \omega_g$ is met. This process, in which signals of two frequencies are combined, is commonly referred to as heterodyne detection. The power transfer is indirectly induced by the deformation of the cavity walls, which leads to the described mode mixing. Additionally, this power transfer can also be caused by the direct interaction of the GW with the electromagnetic field loaded in the cavity. However, this effect—known as the inverse Gertsenshtein effect \cite{gertsensthein,osti_4598733}—is subdominant in the frequency range considered in this work and will not be explicitly addressed.

The idea to detect GWs with superconducting radio-frequency (SRF) cavities dates back to the 1970’s. At the beginning of the 70’s one of the first theoretical concepts explored the Gertsenshtein effect for detection and did not use the heterodyne frequency conversion \cite{Braginskii:1971, Braginskii:1973vm}. Towards the end of the 70’s Pegoraro et al. \cite{Pegoraro:1978gv, Pegoraro:1977uv} and Caves \cite{Caves:1979kq} published papers which proposed the heterodyne detection and the mechanical interaction of GWs with the cavity wall. In the 80’s Reece et al. \cite{Reece:1984gv, Reece:1986zb} started an experimental R\&D programme which was based on the configuration proposed by Pegoraro. This work was based on pillbox cavities and the excited mode was measured through the reflection of the input ports. It was shown that small (order of \SI{e-17}{\centi\metre}) harmonic displacements were detectable with such a superconducting parametric converter. 
 
\begin{figure}[!htbp]
	\centering
		 
  \includegraphics[width=0.8\columnwidth]{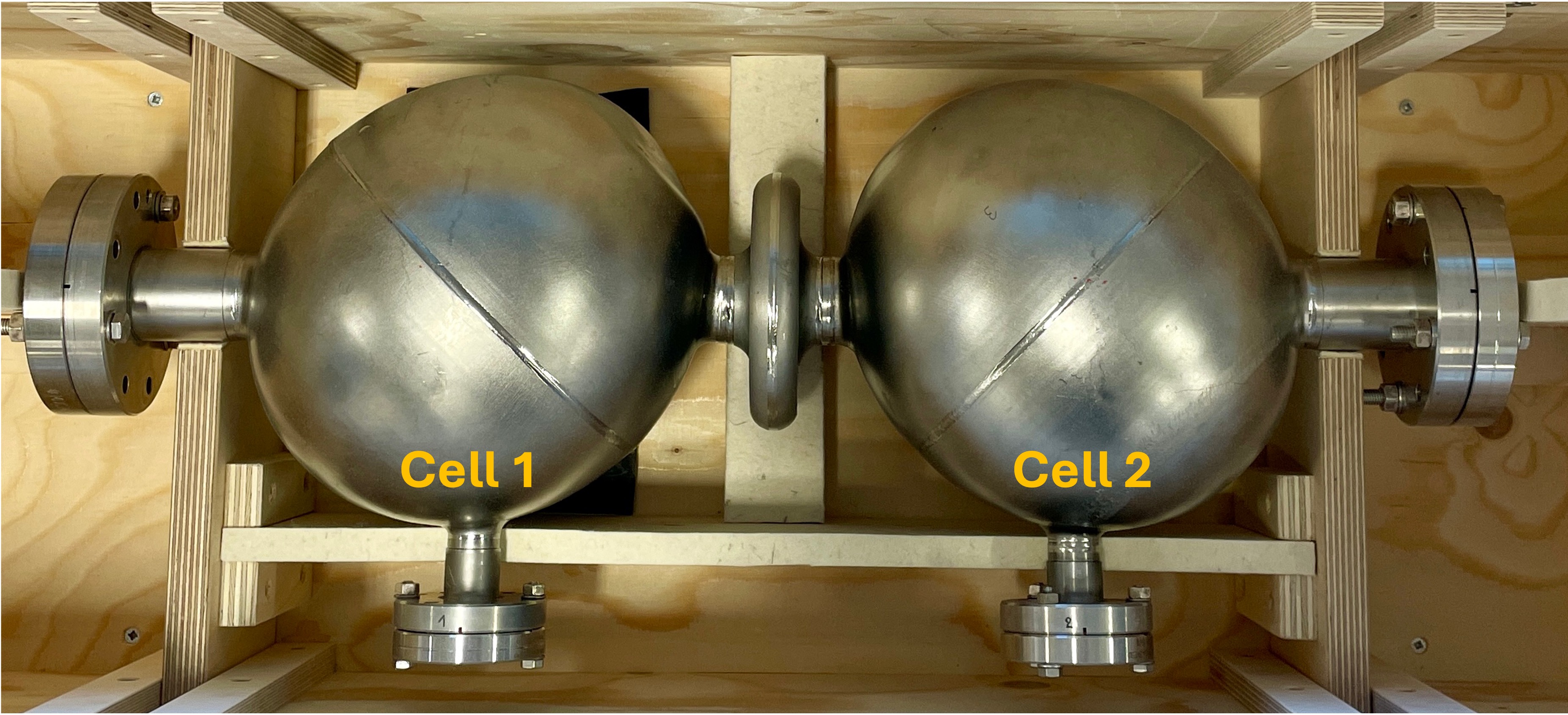}
	\caption{The {\tt PACO-2GHz-variable} prototype niobium spherical cavity. In the picture we highlight which cell we refer as cell 1 vs cell 2.}    	
	\label{fig:MAGO}
\end{figure}

This detection concept was further developed starting at the end of the 90’s within the MAGO proposal with the goal for a scaled-up experiment with 500 MHz cavities as a CERN-INFN collaboration \cite{Ballantini:2003nt, Ballantini:2004wd, Ballantini:2005am}. Since this proposal stems from the time before the discovery of GWs, the aim was to reach frequencies in the lower kHz range which have sensitivity to astrophysical sources. Although the final project was not funded, three SRF niobium cavities were built during the R\&D activities. The first cavity (a pill-box cavity) was used as a proof-of-principle experiment, which demonstrated the working principle and the development of an RF system to drive and read out the cavity with the necessary precision~\cite{Ballantini:2003nt, Ballantini:2004wd, Ballantini:2005am}. The second prototype cavity had two spherical cells with fixed coupling. The third cavity, shown in figure~\ref{fig:MAGO}, was a spherical 2-cell cavity (denoted {\tt PACO-2GHz-variable}) with an optimised geometry and a tunable coupling cell to change the coupling between the cells and so the frequency difference between the two modes. This cavity was never treated nor tested. 

The heterodyne detection with microwave cavities is especially suitable in the higher frequency range kHz to GHz, outside of the LIGO/Virgo detection band, a vast region of frequency range not yet sufficiently explored. An overview of recent experimental efforts in the high-frequency range is summarised in \cite{Aggarwal:2020olq}. The concept was resurrected in recent work~\cite{Berlin:2023grv}, which provided an enhanced analysis of various noise sources and an improved estimate of the resulting sensitivity for a cavity similar to the {\tt PACO-2GHz-variable} cavity. It was found that the noise-equivalent strain power spectral density (PSD) could reach $S_h \sim 10^{-21}/\sqrt{\text{Hz}}$ cross frequencies extending up to the GHz range, approaching the typical signal strengths expected from phenomena such as black hole superradiance. Additionally, by overcoupling to the signal mode of the cavity, the experiment can operate in a broadband mode, allowing two to three decades of frequency to be covered in a single measurement with sensitivities better than $S_h \sim 10^{-18}/\sqrt{\text{Hz}}$.

Because of these promising projections it was decided to continue the R\&D efforts with the {\tt PACO-2GHz-variable} cavity at the University of Hamburg, Deutsches Elektronen-Synchrotron (DESY) and at the Superconducting Quantum Materials and Systems (SQMS) Center at Fermi National Accelerator Laboratory (Fermilab) with measurements at room and at cryogenic temperatures. This paper summarises all of the work which was conducted before the cryogenic RF tests. These include a mechanical survey of the cavity and its room temperature plastic tuning. In addition, it describes the measurements and simulations of the mechanical resonances and electromagnetic properties, which are crucial ingredients for estimating the gravitational wave coupling properties of the specific cavity under investigation.

\section{Mechanical survey}
\label{sec:mechanical}
Since the cavity was fabricated over 20 years ago, limited information was available, and it exhibited significant deviations from the nominal geometry. Therefore, the first crucial step was to conduct a comprehensive survey of both the cavity’s geometry and its wall thickness. This data is essential not only for simulating the mechanical eigenmodes but also for accurately modelling the electromagnetic eigenmodes, as they oscillate within the volume defined by the cavity's geometry. Only with this detailed information can we validate our models and accurately predict the experiment’s sensitivity.

\subsection{Cavity metrology}
\label{sec:geo}
\begin{figure}[!htbp]
	\centering
		\includegraphics[width=0.50\columnwidth]{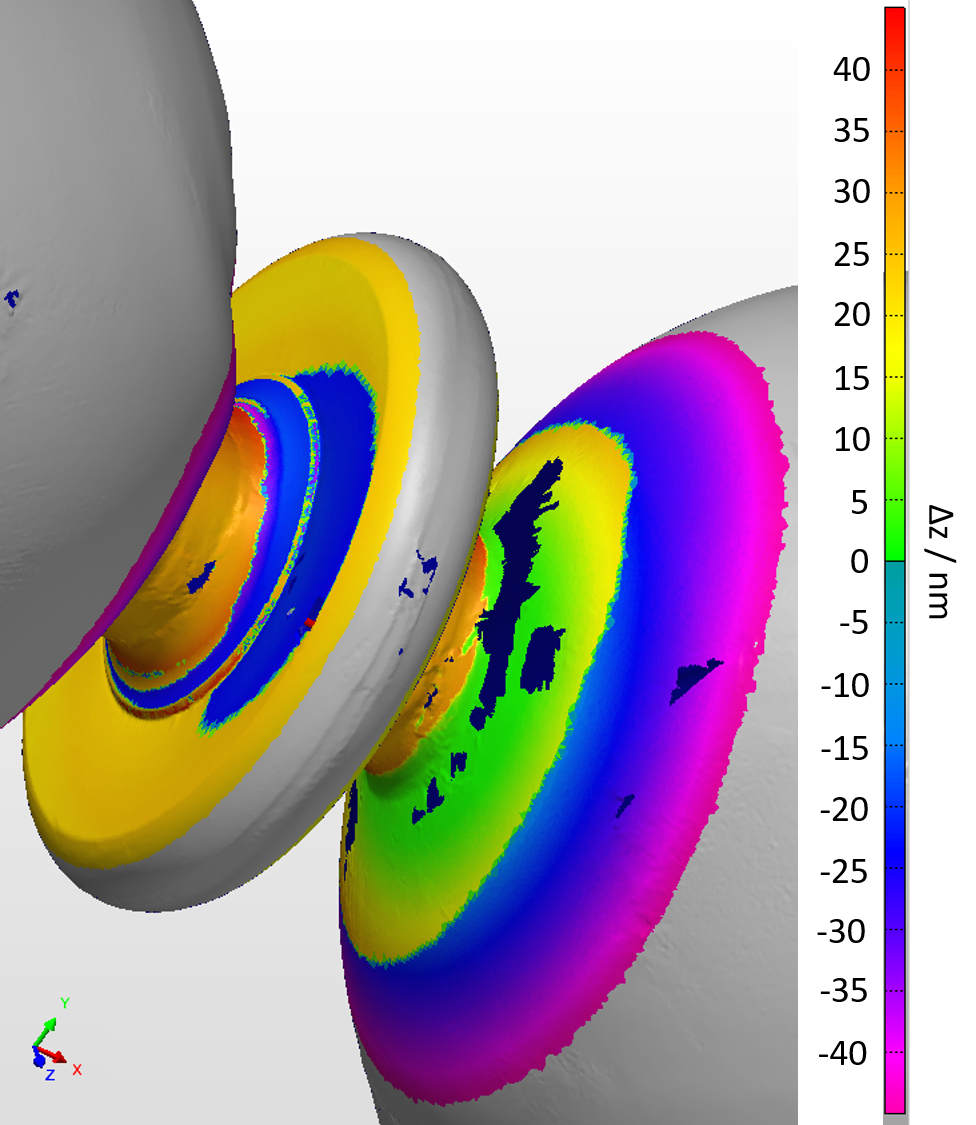}
	\caption{3D Scan of the cavity, focused on the coupling cell. The colour code describes the deviation from the nominal cavity geometry. The dent in the tunable cell is clearly visible by the blue area.}    
	\label{fig:TunableCell}
\end{figure}

\begin{figure}[!htbp]
	\centering
		\includegraphics[width=1.00\columnwidth]{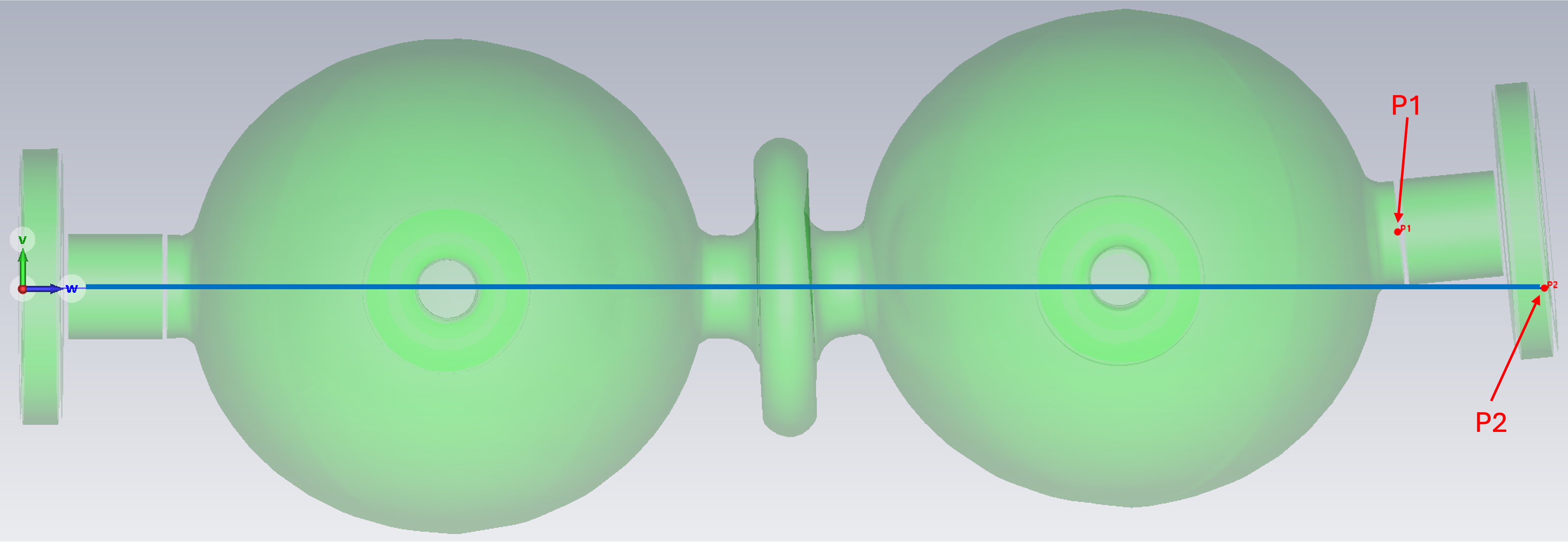}
	\caption{3D visualisation of the cavity. The view is towards the two flanges at the side of each cell. The bending is mostly into one plane and clearly visible. The point $P_1$ on the rotational axis of the tube and the point $P_2$ on the nominal rotational axis of the cavity, highlighted on the right side of the cavity, are nominally expected to sit on the same plane, but are misaligned by 2.3~cm in the vertical plane. For reference, the cavity length, from flange to flange, is approximately 65~cm.}    	
	\label{fig:Shape}
\end{figure}

The cavity was surveyed using a Hexagon Metrology\texttrademark \, 7-axis portable measuring arm equipped with a RS6-Laserscanner. The maximum error for unidirectional length measurements of the system is 0.03\,mm. The inspection showed three severe deviations from the nominal geometry. First, in one of the cells, a large toroidal-shaped dent was found around the rotational axis of the prolate spheroid. After optical inspection, it was found that this dent was also seen inside the cavity. Second, a dent on the so-called \textit{coupling cell}, which is the small cell in the centre, was found, see figure~\ref{fig:TunableCell}.
This dent was aligned with a severe bending of the cavity, starting at the centre, of approximately \ang{6}, shown in figure~\ref{fig:Shape}, which is the third strong deviation of the scanned geometry from the nominal design. 
The cavity shape was scanned using the high-speed laser scanner, yielding a 3D polygon model. By surface reconstruction as a part of the reverse-engineering-process, this point cloud was translated into a fully usable STEP file. This STEP file is the basis for all future simulations, unless it is mentioned that the nominal geometry is used instead. 

\subsection{Wall thickness measurement}
\label{sec:wall}
Another missing piece of information was the thickness of the cavity walls. To obtain this data, a series of 48 measurement points were systematically distributed on the cavity. At an angle of \ang{90} apart, 10 measurement points were taken per cell, 
the wall thickness of the coupling cell was also measured with 8 additional data points. The measurement itself was done using the ultrasonic wall thickness measurement device 38DL PLUS from the company Evident Europe\texttrademark\:using the probe V260-SM with a test frequency 10 MHz. The probe head was connected to a so-called spring loaded holder which ensures a repeat accuracy due to a perpendicular fixed position of the probe on rounded shapes during the measurement. The resolution of the device is \SI{10}{\micro\metre}, but the dominant uncertainty arises from the reproducibility of the positioning of the probe head with respect the surface. This systematic uncertainty can be estimated from repeated measurements at the same spot and is found to be on the order of \SI{50}{\micro\metre}.
For the two spherical cells of the MAGO cavity, a highly non-uniform wall thickness was found. The average wall thickness for cell 1 was found to be $(1.84\pm0.05)$\,mm and for cell 2 to be $(1.89\pm0.05)$\,mm. Point to point variations on the cells of \SI{100}{\micro\metre} was observed, and the welding seams were thicker, as expected. The wall thickness of all the tubes were found to be $(1.91\pm0.02)$\,mm.  

\begin{figure}[!htbp]
	\centering
		\includegraphics[width=0.50\columnwidth]{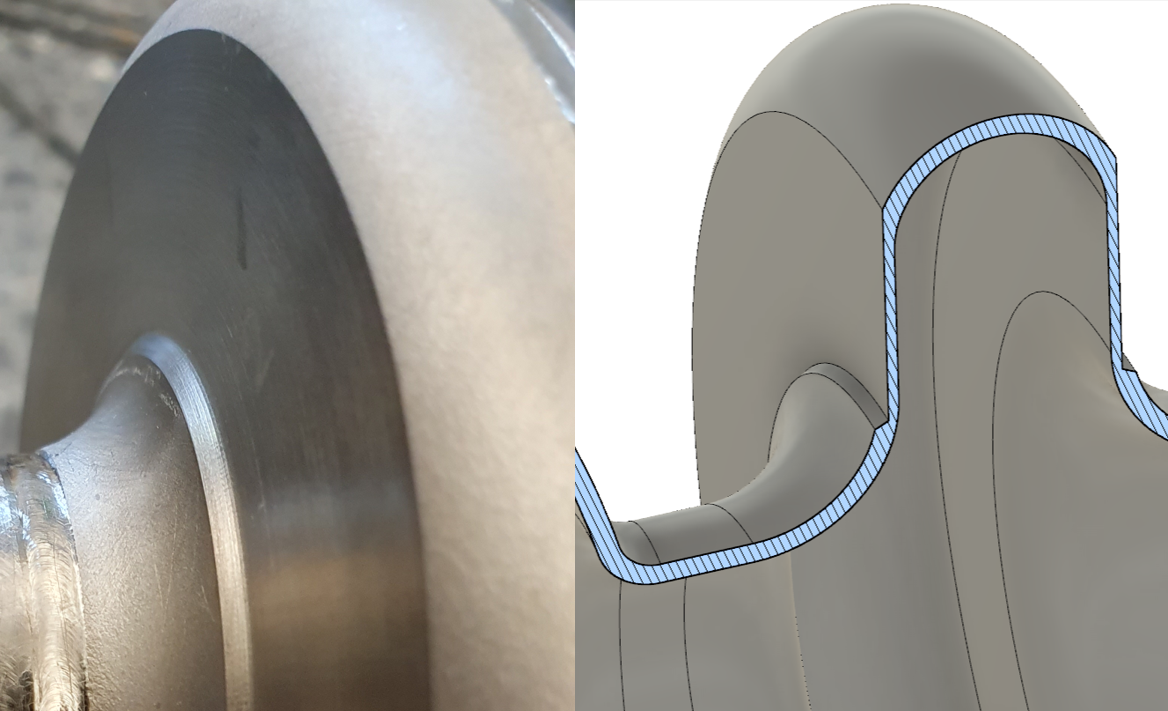}
	\caption{Left: Close up image of the coupling cell. The shiny surface was turned after deepdrawing the niobium sheet. Right: a 3D cut of the turning on the wall thickness showing the impact of the turning. The step seen closer to the centre after turning was on the order of 0.7-1\,mm.}    	
	\label{fig:WallThickness2}
\end{figure}

Another observation is shown in figure~\ref{fig:WallThickness2}.
The coupling cell was dedicated as such -- a mechanically tunable cell to squeeze the cavity longitudinally, and hence bring the two RF cells closer to each other. This would increase the overlap of the RF fields within the cells, and hence increase the cell-to-cell coupling $k_{cc}$. To minimise the force needed to tune the cavity, and to assure that the deformation takes place at the coupling cell and not elsewhere on the cavity, the wall thickness was intentionally reduced. The step shown in figure~\ref{fig:WallThickness2} is on the order of 0.7-1\,mm. The wall thickness measurement just before the step showed an average wall thickness of $(1.0\pm0.1)$\,mm. This is in agreement with the observed wall thicknesses of the RF cells corrected by the estimated removal of the turning. 
\subsection{Mechanical resonances}
\label{sec:resonances}
The indirect detection of high-frequency gravitational waves relies heavily on the mechanical behaviour of the superconducting cavity. Mechanical vibrations, induced by the passing gravitational wave, modulate the RF eigenmodes of the cavity, allowing their reaction to be measured and analysed. For this reason, a study of the mechanical resonances of the MAGO cavity has been conducted.

The technical drawings of the MAGO prototype cavity with the tunable coupling cell were not available, however, they were for the fixed-coupling sibling prototype (denoted {\tt PACO-2GHz-fixed}). Combined with a half recoverable STEP file from the MAGO collaboration of the cavity in use, we constructed a model for the full cavity geometry. A complete and thorough model is important to obtain accurate results in the mechanical  and electromagnetic simulations and also for further development of the low-level RF (LLRF) system. Throughout this work we will refer to this reconstructed geometry as the `nominal' model. 

Since the nominal geometry does not capture all the geometric details of the existing MAGO cavity, we also utilized STEP files obtained from the metrology measurements. These files represented the cavity as a hollow shell without any wall thickness. While the cavity walls are known to vary in thickness, as discussed earlier, we approximated a uniform wall thickness of 1.85 mm. We will refer to this geometry, based on the scan data, as the `scanned' model.

The material properties used for the simulation are summarised in table \ref{tab:Nbmaterial} and the software used for finite element method simulations was COMSOL\texttrademark.
\begin{table}[htbp]
\caption{\label{tab:Nbmaterial} Material properties used for Nb.}
$$
\begin{tabular}{c c c c c }
 \hline
 \multicolumn{1}{c}{Temperature} &\multicolumn{1}{c}{Grain size} &\multicolumn{1}{c}{Density} &\multicolumn{1}{c}{Young's modulus} &\multicolumn{1}{c}{Poisson's ratio $\nu$ }  \\
\hline		
	293\,K & $45-90\,\mu m$  & 8570\,kg$\cdot$ m\textsuperscript{-3} & 106\,GPa & 0.4 \\
\hline
		\end{tabular}
$$		
\end{table}

\label{sec:res_sim}
\begin{figure}[!htbp]
	\centering
		\includegraphics[width=0.7\columnwidth]{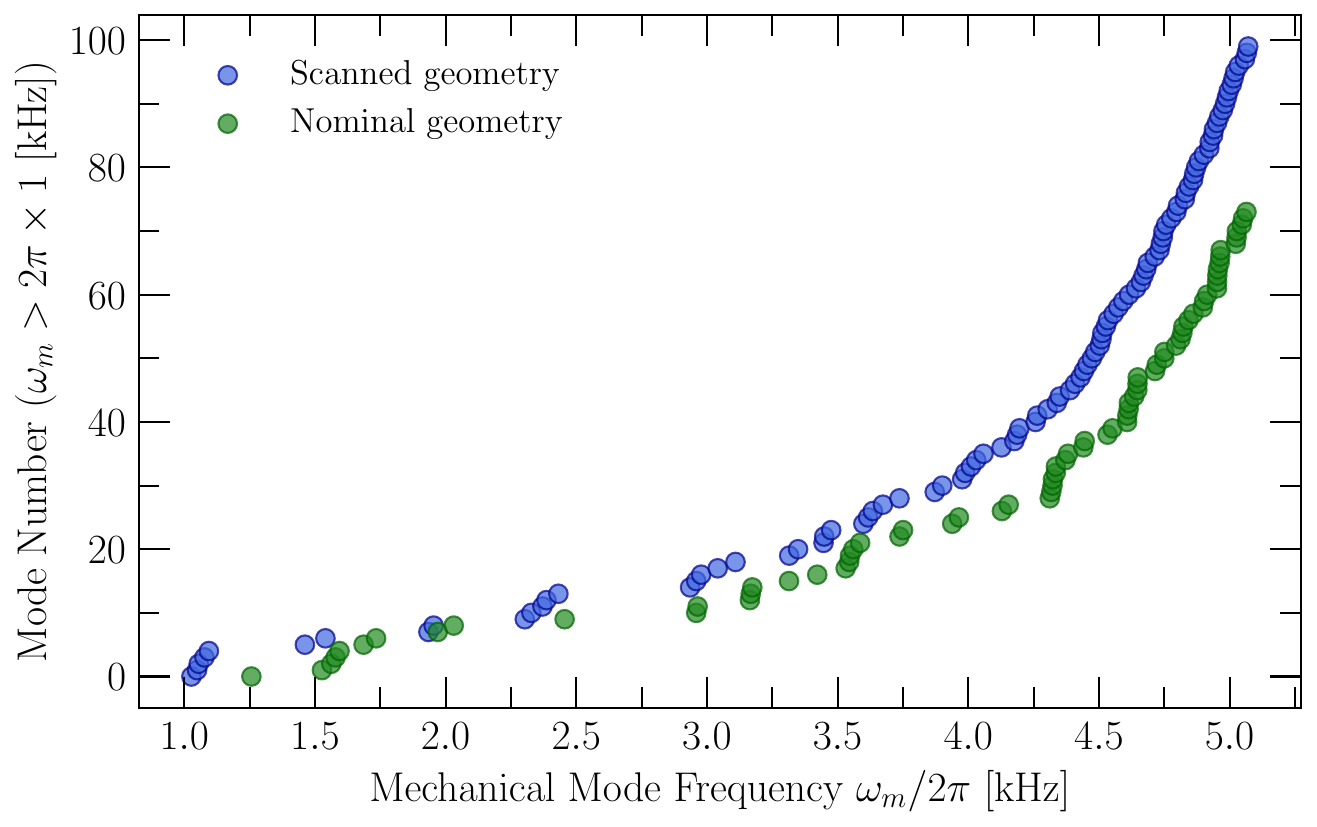}
	\caption{Cumulative number of mechanical eigenmodes for the nominal (green) and the scanned (blue) cavity geometry above $\omega_m > 2\pi\times \SI{1}{kHz}$. A significant increase of total number and a shift of the frequencies of eigenmodes is observable for the scanned cavity model.}    	
	\label{fig:Nommodes}
\end{figure}

A first impression of the complexity of this study is shown in figure \ref{fig:Nommodes}, which shows a comparison of the amount of eigenmodes of the nominal and scanned cavity geometry.
Besides a shift of the eigenfrequencies and an increase of the eigenmodes, the deformation patterns of the eigenmodes became more complex due to the bending and dent of cell 1. Hence, it was not straightforward to identify the lowest lying quadrupole mode, which is the most sensitive mode for a gravitational wave.

Measurements were carried out using the FPS3010 optical interferometer from attocube\texttrademark. 
The cavity was mounted on a  Rose\&Krieger\texttrademark\,structure and placed on an optical table to minimise mechanical noise from the surrounding laboratory environment. Additionally, the support structure of the interferometer was positioned on a concrete block, which further helped reduce noise levels and vibrations affecting the support structure itself. 

No agreement has been found between the simulation of the scanned geometry and the measurements for the MAGO cavity, which is likely due to the oversimplified model, which uses a first-order approximation assuming a homogeneous wall thickness. This approximation does not accurately reflect the complex reality of the cavity's structure, leading to the observed discrepancies. In addition, the toroidal-shaped dent in one of the cells distorts the mechanical eigenmodes significantly. This discrepancy is currently under investigation with benchmark measurements of a simpler cavity geometry.

\section{Electromagnetic properties}
\label{sec:em}

\begin{figure}[!h]
    \centering
    \includegraphics[width=1.00\columnwidth]{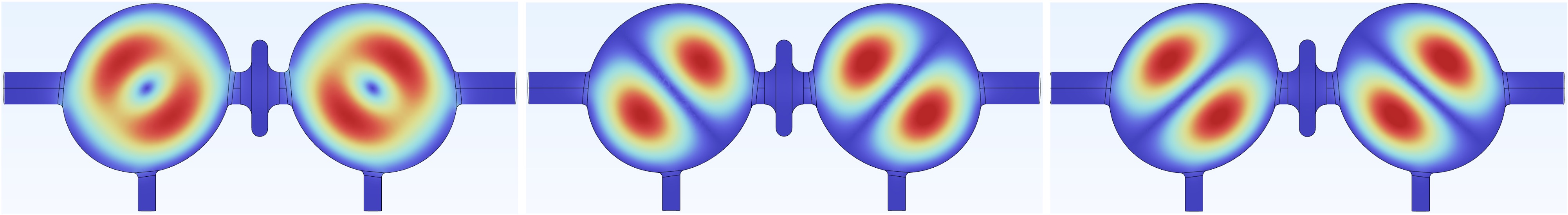}
    \caption{Electric field profile of the three  $\text{TE}_{011}$ pairs of the quasi-degenerate symmetric and anti-symmetric modes. According to the electromagnetic simulations conducted on the nominal geometry, for each pair the separation between the symmetric and anti-symmetric modes is equal to few KHz (from 7 to 21\,KHz), while the pairs are separated by $>\SI{1}{\mega\hertz}$ from each other. The modes of interest for this search are the pair on the right.}    	
    \label{fig:TE011modes}
\end{figure}

The $\text{TE}_{011}$ mode is the electromagnetic mode of interest for the GWs search with the MAGO prototype cavity. When coupled together, the two nominally identical spherical cells give rise to a family of three pairs of $\text{TE}_{011}$ quasi-degenerate symmetric and anti-symmetric modes. Figure~\ref{fig:TE011modes} shows the electric field profiles for the three $\text{TE}_{011}$ pairs. For each pair, two modes exist with identical field patterns and with a $0$ or $\pi$ phase difference. For this search, we are interested in the third pair of $\text{TE}_{011}$, and we will refer to the two modes with $0$ or $\pi$ phase difference as symmetric and anti-symmetric respectively. The frequency splitting between the symmetric and anti-symmetric modes can be controlled by tuning the coupling cell.

\subsection{Simulation of nominal and scanned geometry}
\label{sec:comp}

\begin{figure}[htbp]
    \centering
    \includegraphics[width=1.00\columnwidth]{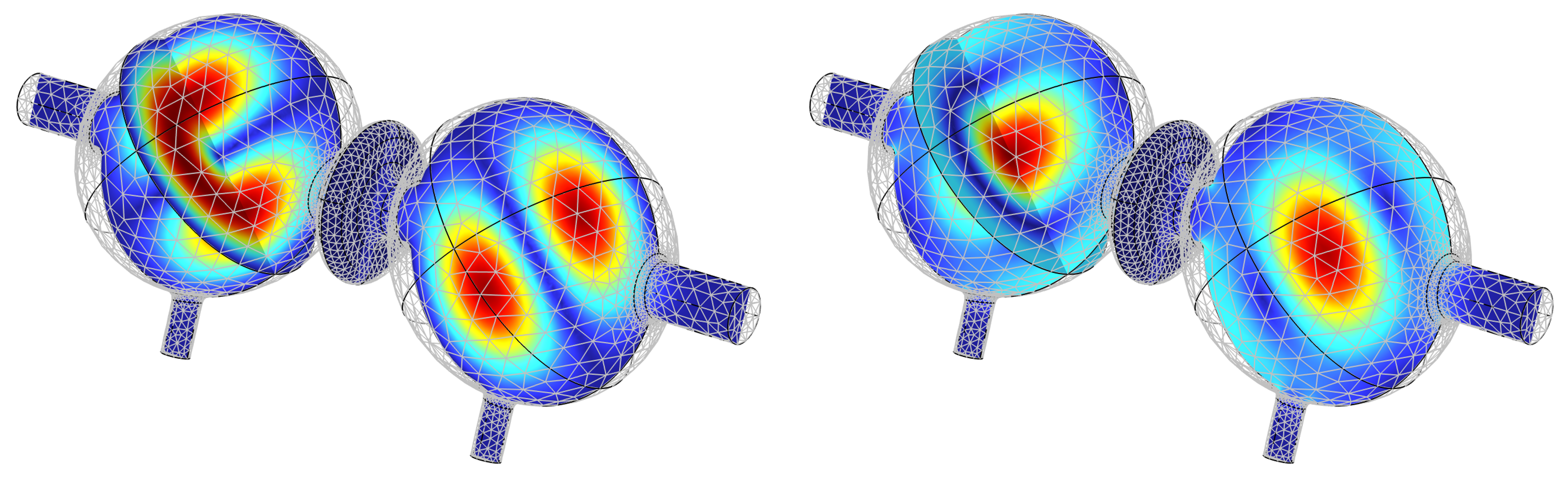}
    \caption{Electric (left) and magnetic (right) field norm of the symmetric and anti-symmetric $\text{TE}_{011}$ mode which only differ in phase and have identical field patterns. The B-field strength in the 
    coupler ports is larger, thus the use of a loop antenna design is planned for cryogenic RF tests.}    	
    \label{fig:EfieldTE011}
\end{figure}

Our simulations based on the nominal cavity geometry yield a resonance frequency of $\omega\simeq2\pi\times\SI{2.1}{GHz}$ for the $\text{TE}_{011}$ modes of interest, and the expected splitting between the symmetric and anti-symmetric modes of $\Delta\omega\sim 2\pi\times\SI{10}{kHz}$. In figure \ref{fig:EfieldTE011} we show the field distribution of the desired $\text{TE}_{011}$ modes. Note that an average value of the thickness over both cells has been used for the wall dimensions, where a thickness difference $\Delta d=\SI{10}{\micro\metre}$ results in an absolute eigenmode frequency change of $\Delta\omega\simeq 2\pi\times\SI{0.2}{\mega\hertz}$ while preserving the splitting distance of symmetric and anti-symmetric mode.

A simulation of the eigenmode spectrum for the scanned geometry revealed that the eigenfrequencies of the $\text{TE}_{011}$ mode in the two cells are separated by more than $\SI{1}{\mega\hertz}$, as plotted in figure \ref{fig:MAGOsimulationVsMeasurement}, and have a poor field flatness with an amplitude ratio on the order of $\mathcal{O}(10^2)$. The geometry of the two spherical cells deviate significantly, causing their respective eigenfrequencies to be too far apart. As a result, the symmetric and anti-symmetric modes are primarily determined by the individual cell frequencies. This was also confirmed with first RF measurements at room temperature, see section \ref{sec:rf}. A theoretical explanation for this behaviour, based on RLC equivalent circuit models, is provided in section \ref{sec:rf}. 

\subsection{RF measurements}
\label{sec:rf}

\begin{figure}[htbp]
    \centering
  
    \includegraphics[width=.8\columnwidth]{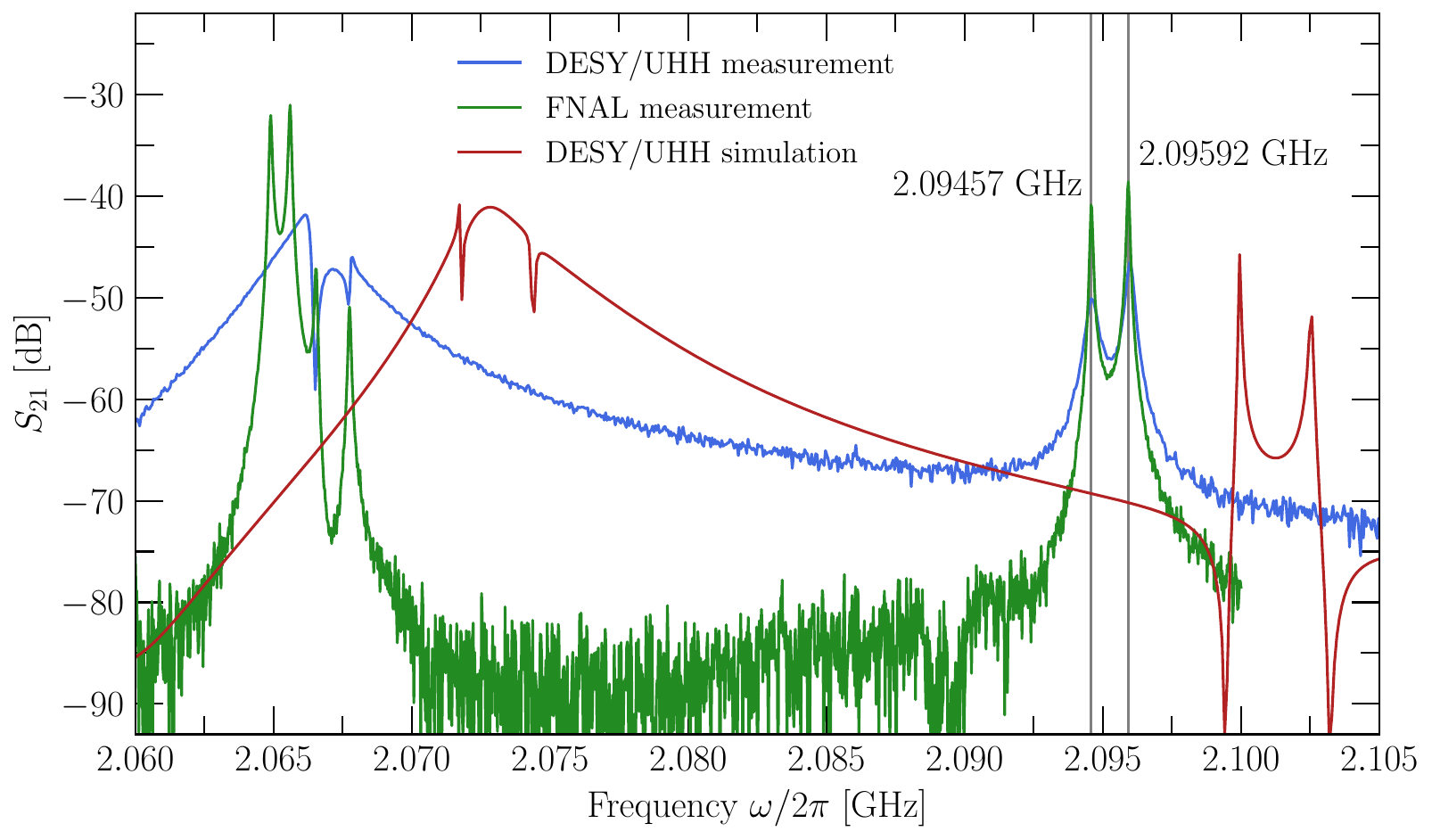}
    \caption{$S_{21}$ measurements and simulation of the cavity at room temperature. Measurements at DESY (blue) have been performed with a pin-antenna which couples to the electric field of the modes, while the measurement at
Fermilab (green) used a loop-antenna which couples to the magnetic field of the modes. The simulations (red) have been performed using the DESY antenna setup, using the scanned geometry and an averaged wall thickness.
    }
    \label{fig:MAGOsimulationVsMeasurement}
\end{figure}
\begin{table}[h]
\centering
\begin{tabular}{|c|c|c|c|}
\hline
DESY/UHH&FNAL&Simulation\\
\hline
\hline
        &2064.9&\\
        &2065.6&\\
        \hline
2066.1   &2066.5&    2071.6\\
2067.9   &2067.8 &    2074.6\\
\hline
2094.6   &2094.6  &  2099.9\\
2096.0   &2096.0   & 2102.5\\
\hline
          
\end{tabular}
\caption{The frequencies of the resonances  of the three distinct polarities
of the $\text{TE}_{011}$ mode from figure \ref{fig:MAGOsimulationVsMeasurement} in MHz.}
\label{table:maxima}
\end{table}

In order to verify the electromagnetic simulations, RF measurements have been carried out at DESY and Fermilab by obtaining an $S$-parameter spectrum characterising the resonances of the system. The results are shown in figure \ref{fig:MAGOsimulationVsMeasurement}.  The $S_{21}$ measurement at DESY used a pin-antenna which couples to the electric field of the modes, while the measurement at Fermilab used a loop-antenna which couples to the magnetic field of the modes. Comparing these $S_{21}$ measurements, the impact of the antenna is obvious at the lower lying $\text{TE}_{011}$ modes. The loop-antenna leads to a better separation of the three distinct polarities of the $\text{TE}_{011}$ mode, where the field distributions are shown in figure \ref{fig:TE011modes}, resulting in different resonance peaks in the $S_{21}$ measurement, see table~\ref{table:maxima} .

Comparing to the simulation, there is a clearly visible constant shift in the spectrum of around 5 MHz. 
As for the simulation of the mechanical resonances, we assume an averaged wall thickness of $\SI{1.85}{mm}$.
However, we find that the effective wall thickness would have to be $\sim \SI{250}{\micro m}$ thinner compared to the measured wall thickness to match the measured spectrum, which results in a sensitivity of $\SI{20}{\kilo\hertz/\micro m}$.

The cavity was designed to have symmetric and anti-symmetric oscillations of the $\text{TE}_{011}$ mode with a frequency spacing of $\mathcal{O}(10)\, \text{kHz}$ \cite{Ballantini:2005am}. However, the spectrum reveals
an apparent frequency spacing of $\mathcal{O}(1)\,\text{MHz}$ instead, see figure \ref{fig:MAGOsimulationVsMeasurement}.
Each of the two cells can be measured individually by either shorting one cell and measuring the other, or by performing an $S_{21}$ measurement using loop antennas inserted into the flanges of the same cell. Both measurement methods show good agreement, and we found that the eigenfrequencies of the individual cells closely matched one of the resonant frequencies of the seemingly coupled system, indicating that the cells have a low coupling coefficient and behave as an uncoupled system at room temperature. In order to understand this behaviour in more detail, we can turn to an equivalent circuit model for the MAGO cavity.

\subsection{Equivalent circuit modelling}
\label{sec:LRC}
\begin{figure}[h]
\centering
\begin{circuitikz}[american]
\draw (0,0) node [transformer](T){} 
(T.west) node{$L_c^{(1)}$}
(T.east) node{$L_c^{(2)}$}



(T.base) node{$L_\text{cpl}$};
\draw (T.A1) to[R, -, l = $R_c^{(1)}$] ++(-1,0)--++(-1,0);
\draw[very thick](T.A1)+(-4,0)node[ocirc]{}--++(-2.05,0)node[ocirc]{};
\draw (T.A2) to[C, -, l = $C_c^{(1)}$] ++(-1,0)--++(-1,0);
\draw[very thick](T.A2)+(-4,0)node[ocirc]{}--++(-2.05,0)node[ocirc]{};

\draw (T.B1) to[R, -, l = $R_c^{(2)}$] ++(1,0)--++(1,0);
\draw[very thick](T.B1)+(4,0)node[ocirc]{}--++(2.05,0)node[ocirc]{};
\draw (T.B2) to[C, -, l = $C_c^{(2)}$] ++(1,0)--++(1,0);
\draw[very thick](T.B2)+(4,0)node[ocirc]{}--++(2.05,0)node[ocirc]{};

\draw (-4.6,1) to [open, l=$Z_\text{in}$] (-4.6,-1);
\draw (3.5,1) to [open, l=$Z_\text{out}$] (3.5,-1);

\end{circuitikz}
\caption{The equivalent RLC circuit for a MAGO-like cavity with one RF port per cavity cell.}
\label{fig:2_port_circuit}
\end{figure}
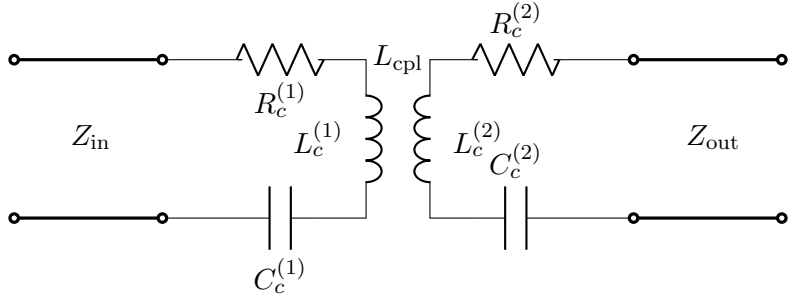
The two coupled cells of the MAGO cavity can be modelled as two inductively coupled RLC circuits with a transmission line connected to an RF port in each cell (figure \ref{fig:2_port_circuit}). We denote the input and output impedance of the ports as $Z_\text{in}$ and $Z_\text{out}$ respectively. The impedance of each cell is given by $Z_c^{(j)}(\omega)=R_c^{(j)}+i\,X_c^{(j)}(\omega)$, where
\begin{equation}
    X_c^{(j)}(\omega)\coloneqq\omega L^{(j)}_c-\frac{1}{\omega C^{(j)}_c},\qquad j=1,2
\end{equation}
and the coupling impedance depends on the inductance $L_\text{cpl}$
\begin{equation}
    Z_\text{cpl}(\omega)\coloneqq i\,X_\text{cpl}(\omega)\coloneqq i\omega L_\text{cpl}\,.
\end{equation}
This allows us to derive the general scattering matrix of our linear network
\begin{align}\label{eq:2_port_S_matrix}
    \mathbf{S}=&\frac{1}{(Z_c^{(1)}+Z_\text{in})(Z_c^{(2)}+Z_\text{out})-Z_\text{cpl}^2}\\ \nonumber
&\times\begin{pmatrix}
(Z_c^{(1)}-Z_\text{in})(Z_c^{(2)}+Z_\text{out})-Z_\text{cpl}^2 & 2\sqrt{Z_\text{in}Z_\text{out}}Z_\text{cpl} \\
2\sqrt{Z_\text{in}Z_\text{out}}Z_\text{cpl} & (Z_c^{(1)}+Z_\text{in})(Z_c^{(2)}-Z_\text{out})-Z_\text{cpl}^2
\end{pmatrix}\,.
\end{align}
If the internal resistances $R_c^{(j)}$, $Z_\text{in}$ and $Z_\text{out}$ are negligible, we can find two separate resonance frequencies of this system from the criterion 
\begin{equation}
    X_c^{(1)} X_c^{(2)}-X_\text{cpl}^2=0\,.
\end{equation}
The solution depends on the coupling strength between the cells $k_{cc}\coloneqq L_\text{cpl}/\sqrt{L_c^{(1)}L_c^{(2)}}$, as well as the difference in the parameters of the cells $\tilde{k}_{12}\coloneqq\frac{\tilde{\omega}_2-\tilde{\omega}_1}{\omega_c}$, where $\tilde{\omega}_{j}=(L_c^{(j)}C_c^{(j)})^{-1/2}$ are the single cell eigenfrequencies and $\omega_c\coloneqq(\tilde{\omega}_2+\tilde{\omega}_1)/2$. Using this, we find for the two eigenfrequencies of the coupled system
\begin{equation}  \label{eq:eigenfrequencies}
\omega_{\pi/0}=
\begin{dcases*}
    \frac{\omega_c}{\sqrt{1\mp k_{cc}}}\quad &\text{for $\tilde{k}_{12}\ll k_{cc}\ll1$} \\
    \tilde{\omega}_{2/1}\quad &\text{for $k_{cc}\ll\tilde{k}_{12}\ll1 $}
\end{dcases*}\,.
\end{equation}
Apart from some exceptions discussed later, the two single cells oscillate \emph{in phase} when driven at $\omega_0$ and \emph{out of phase} (i.e.\ with phase shift $\pi$) when driven at $\omega_\pi$.
We see that for two similar cavity cells with $k_{cc}\gg\tilde{k}_{12}$, the two eigenfrequencies depend on the coupling $k_{cc}$. In this regime, we can also turn this relation around in order to estimate the coupling from the measured eigenfrequencies
\begin{equation}
    k_{cc}=\frac{\omega_\pi-\omega_0}{(\omega_\pi+\omega_0)/2}\,.
    \label{eq:kcc}
\end{equation}
However, if the coupling strength drops even further, or the two single cells have significantly different parameters, the eigenfrequencies of the coupled system converge to the eigenfrequencies of the single cavity cells. This also means that the value of $k_{cc}$ can no longer be inferred from $\omega_0$ and $\omega_\pi$ alone. As both the spectrum of the single cavity cells and of the coupled system were measured, we can conclude that the MAGO cavity was measured in the regime $k_{cc}\ll\tilde{k}_{12}\ll1$.

\begin{figure}[!htbp]
	\centering
		\includegraphics[width=0.7\columnwidth]{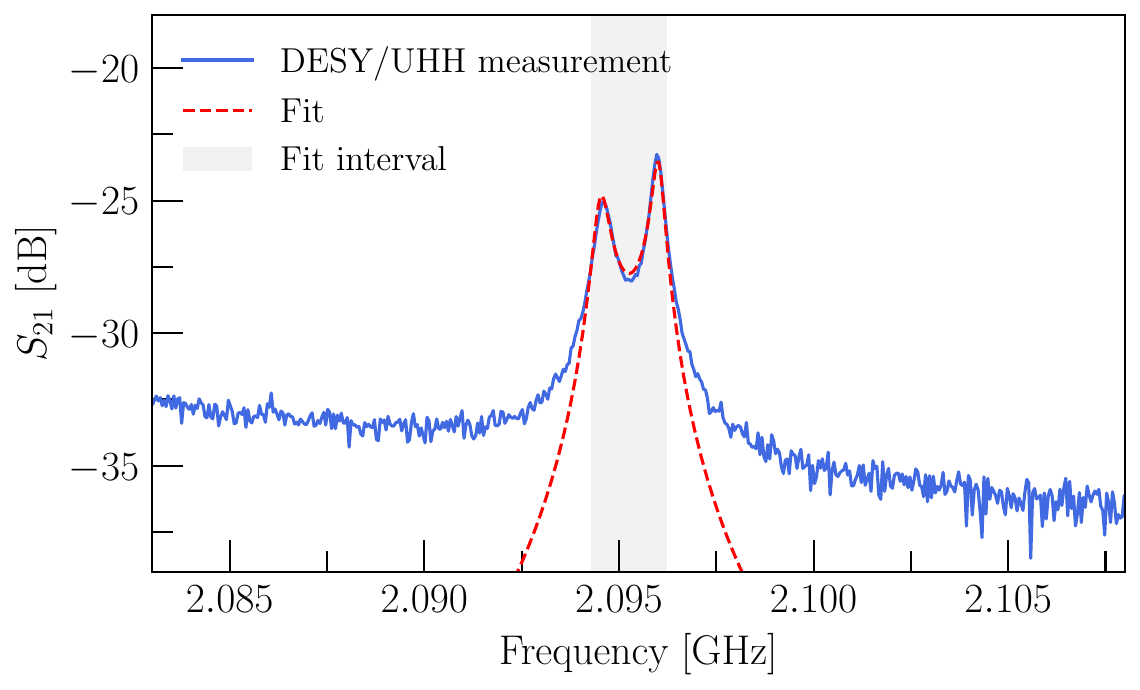}
	\caption{The $S_{21}$ parameter from equation \eqref{eq:2_port_S_matrix} fit to the measured data of DESY from figure~\ref{fig:MAGOsimulationVsMeasurement}. The fitted parameters are $\tilde{\omega}_1=2.09457$ GHz, $\tilde{\omega}_2=2.09597$ GHz, $k_{cc}=2\cdot 10^{-4}$, $Q_1=5\cdot 10^3$, $Q_2=8\cdot 10^3$. The $k_{cc}$ for the other measurements are shown in figure \ref{fig:ampratios}. For $R_c^{(1)}\gg Z_\text{in}$ and $R^{(2)}_c\gg Z_\text{out}$ all resistances only appear in the overall factor $\sqrt{Z_\text{in}Z_\text{out}}/\sqrt{R_c^{(1)}R_c^{(2)}}$, which takes the value $0.013$.}    	
	\label{fig:RF_Spectrum_Fit}
\end{figure}

More details can be inferred by fitting the $S_{21}$ parameter from Eq. \eqref{eq:2_port_S_matrix} to the measurements, which are illustrated in figure \ref{fig:RF_Spectrum_Fit}. The frequency spacing of the resonance peaks is most consistent with $k_{cc}\approx10^{-4}$ and the different peak heights are well explained by different quality factors in both cells. This is a significantly weaker value than the cell-to-cell coupling in superconducting TESLA cavities, which typically have $k_{cc}\sim 10^{-2}$ \cite{PhysRevSTAB.3.092001}.

\begin{figure}[!htbp]
	\centering
		\includegraphics[width=.7\columnwidth]{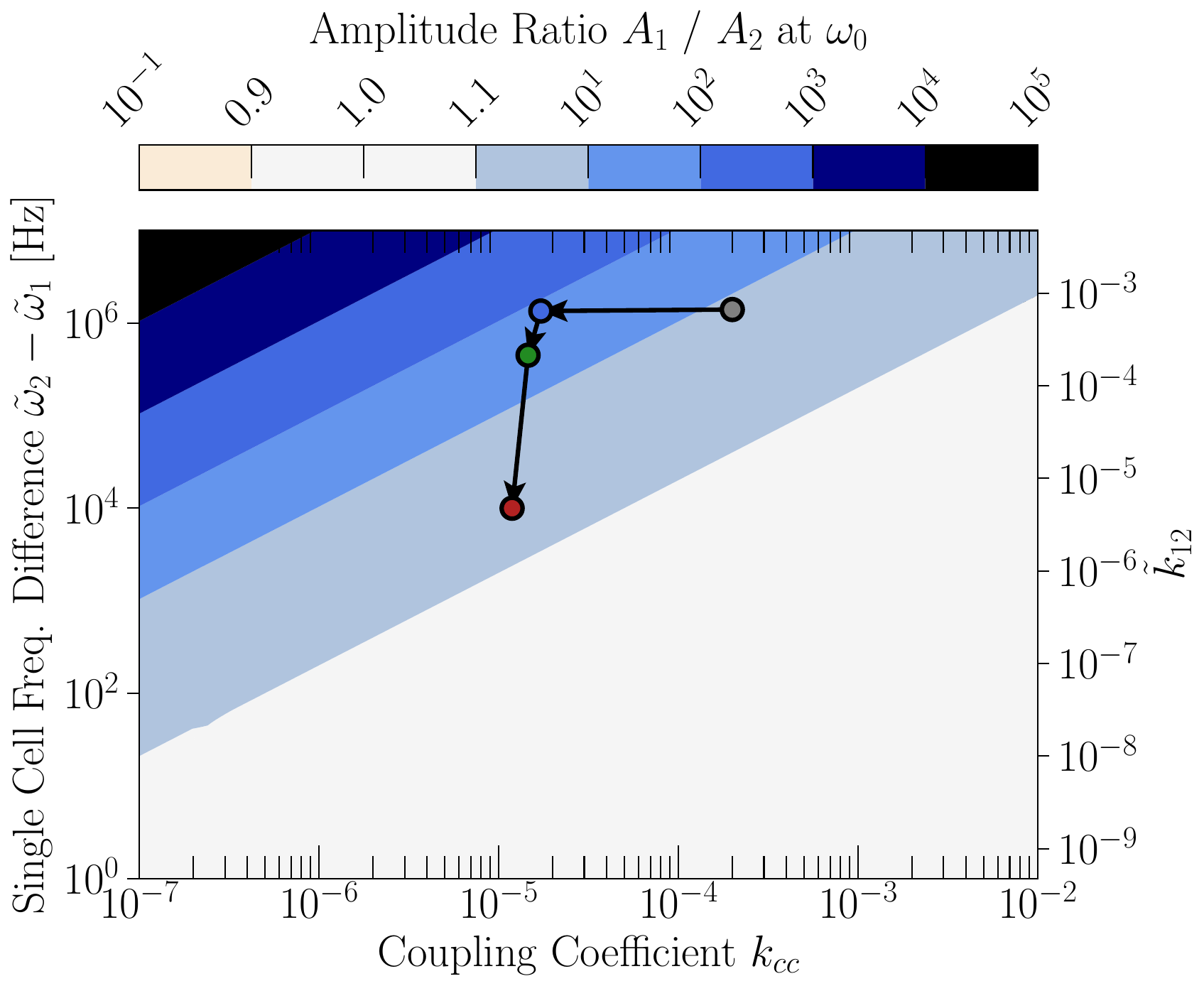}
	\caption{The amplitude ratio according to Eq.\ \eqref{eq:ampratio_both} of two coupled circuits driven symmetrically on \emph{both} sides for  different values of $k_{cc}$ and $\tilde{k}_{12}$. The ratio is evaluated at the lower resonance frequency $\omega_0$. We have chosen $\tilde{\omega}_1=2.1$ GHz and $Q_1=Q_2=10^{10}$. The point marked in grey corresponds to the state of the cavity when it first arrived at DESY and is taken from the fit in figure~\ref{fig:RF_Spectrum_Fit}. The three subsequent points are obtained from the three tuning steps in figure~\ref{fig:FinalSpectra} where the cavity was first straightened and then tuned with a chain-tuner (see section \ref{sec:tuning}). }   	
	\label{fig:ampratios}.
\end{figure}

The two coupled cells having significantly different impedances impacts the gravitational wave sensitivity of the resonator in three separate ways. First, the frequency spacing of the pump and signal mode changes according to Eq. \eqref{eq:eigenfrequencies}. Second, the two cells may not oscillate with the same field amplitude. Third, the phase shift between signal and pump mode may take on values other than $0$ or $\pi$.
The amplitude ratio for the cavity driven at the input port is given by 
\begin{equation}
    \frac{A_1}{A_2}=\left|\frac{Z_c^{(2)}(\omega)+Z_\text{out}}{Z_\text{cpl}(\omega)}\right|\qquad(\text{One Sided})\,.
\end{equation}
Another interesting mode of operation is when both sides of the coupled cavity are driven at the same time. In this case, the amplitude ratio becomes
\begin{equation}\label{eq:ampratio_both}
    \frac{A_1}{A_2}=\left|\frac{Z_c^{(2)}(\omega)+Z_\text{out}\mp Z_\text{cpl}(\omega)}{Z_c^{(1)}(\omega)+Z_\text{in}\mp Z_\text{cpl}(\omega)}\right|\qquad(\text{Two Sided})\,,
\end{equation}
where the upper (lower) sign corresponds to both ports being driven (anti-) symmetrically. Figure~\ref{fig:ampratios} shows the amplitude ratios for a coupled cavity driven symmetrically on both sides for different couplings and values of $\tilde{k}_{12}$. A very similar picture emerges if only one cell is driven. Parametrically, we find $A_1/A_2\sim 1+\tilde{k}_{12}/k_{cc}$ i.e.\ the weaker the coupling between the cells is, the more sensitive is the amplitude ratio to a difference in the two cells. Amplitude ratios $\neq1$ affect the gravitational wave sensitivity of the coupled cavity in two ways. First, an amplitude difference only allows the amplitude in one cell to be operated at the maximal (quench limited) field values. This lowers the maximum power at which the cavity can operate and in our case the GW sensitivity. Second, the mechanical-electromagnetic coupling $C_{01}^m$ multiplies the pump and signal fields. Since the lower field amplitude usually switches sides between $\omega_0$ and $\omega_\pi$, this lowers our coupling coefficient by a factor of about $A_2/A_1$.

Arbitrary phase shifts $\neq0,\,\pi$ can lower the gravitational wave sensitivity as well since it is unlikely to find mechanical vibrations that match the phase shift and couple well to the EM mode. However, it turns out that such arbitrary phase shifts only occur for quality factors $\lesssim10^6$ and can also be eliminated by driving both cells of the cavity simultaneously.
Therefore, arbitrary phase shifts do not affect the GW sensitivity of a coupled SRF cavity.

Overall, the study of the equivalent circuit leads us to two conclusions. First, whenever two weakly coupled cavity cells need to have the same field amplitude in both cells, there are stringent requirements on how similar both cavity cells need to be manufactured. Second, the MAGO prototype cavity as we received it did not meet these requirements and needed to be tuned in order to restore the gravitational wave sensitivity it was designed to have.

\section{Cavity tuning}
\label{sec:tuning}
In order to improve the merging of the two cells into a coupled system, for the given cell-to-cell coupling $k_{cc}$, it was necessary to plastically deform the individual cells and therefore tune their RF eigenfrequencies. This would then decrease the observed difference between the two eigenfrequencies from the order of $\mathcal{O}(\text{MHz})$ down to the intended $\mathcal{O}(\text{kHz})$. It should be noted that such a weak cell-to-cell coupling $k_{cc}$ is inherent to the detection scheme. As the difference of the eigenmodes is designed to be $\mathcal{O}(\text{10\,kHz})$, and the frequency of the $\text{TE}_{011}$ mode is $\mathcal{O}(\text{2\,GHz})$, equation \ref{eq:kcc} results in a  cell-to-cell coupling $k_{cc}$ of $\sim 10^{-5}$.

\begin{figure}[!htbp]
	\centering
		\includegraphics[width=0.85\columnwidth]{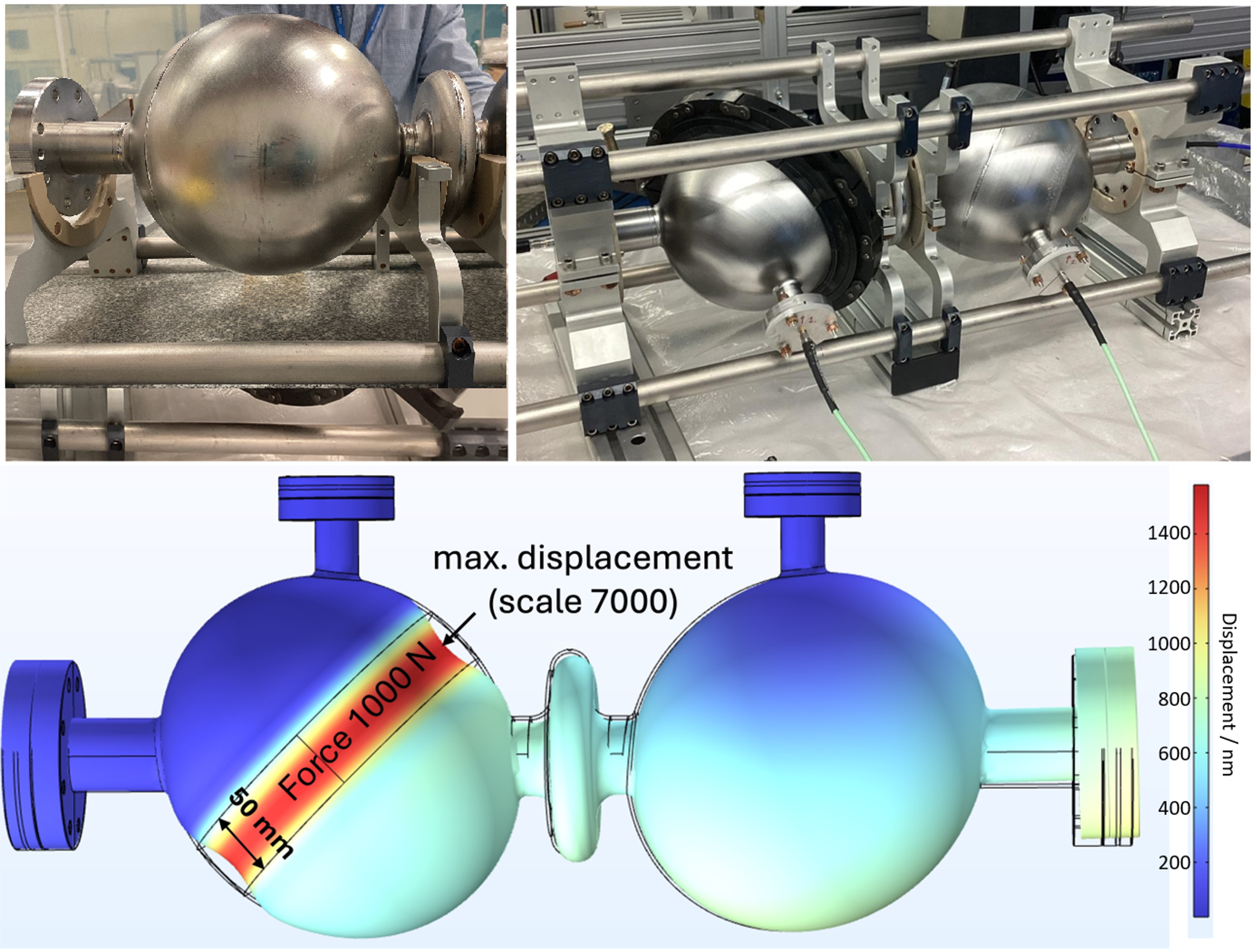}
	\caption{Top left shows the cavity as installed on the bottom half of the frame, prior to straightening: the left flange does not rest on the frame due to the cavity bending. The top right shows the chain tuner installed around the minor axis of cell 1.  Bottom: Deformation study on the RF cell, simulating the effect of the chain-tuner. The strategy was to squeeze a waist around the minor axis of the ellipsoid.}
	\label{fig:TuningSim}
\end{figure}

First, the bend in the cavity, shown in figure~\ref{fig:Shape}, was eliminated in order to facilitate preparation steps of the inner cavity surface such as a buffered chemical polishing (BCP) which removed $\approx5\,\mu$m from the cavity inner surface, and high pressure rinsing in preparation for the cold RF test. These activities will be the subject of a future publication, covering also the results of the RF cryogenic tests. At Fermilab, the cavity was constrained in a frame designed to support the cavity during the RF cold tests. Removing one of the external support spiders, it was possible to delicately push on the cavity flange until the bent was removed and the cavity was plastically straightened, see top left of figure~\ref{fig:TuningSim}. This step was conducted after a heat treatment of twenty four hours at 600\,\textsuperscript{o}C under vacuum, to anneal the material and relieve the fabrication stress. Comparative measurements of the $S_{ij}$ parameters of the cavity before and after each step showed no measurable change in the eigenfrequencies. The cavity was assembled with four small loop antennas for this step, and the following tuning, to monitor the frequency changes via a 4-port vector network analyser.

Cavity production for projects, such as the European XFEL,  showed that a cell shape accuracy on the level of 1\,\% can be expected \cite{Singer2016}. However, the cavity production of the European XFEL was standardised and well monitored, while the MAGO cavity was fabricated as a prototype, hence further deviation can be expected. As a result, the radii of the two cells could differ by about $\pm2$\,mm or even more.
For the tuning, it was planned to plastically deform cell 1, the cell with the lower eigenfrequency, by squeezing the minor axis, since the TE\textsubscript{011} mode is the most sensitive to shape deviations in this region, see figure~\ref{fig:TuningSim}. 

To prepare for the room temperature plastic tuning of the cavity, several displacement and RF simulations were carried out. The goal was to estimate the order of magnitude for the sensitivity coefficient of the frequency change as a function of the cell deformation. Figure~\ref{fig:TuningSim} depicts the so-called \textit{chain-tuner} acting on the cavity circumference, with a contact width of \SI{50}{\milli\metre}. The obtained sensitivity coefficient is equal to 4.2\,MHz/mm.

The chain-tuner was installed around the minor axis of the cell, see figure~\ref{fig:TuningSim}. To apply a force, a spindle was installed between two links, where each rotation of the spindle tightened the chain-tuner and increased the applied force.
An approach of alternating tightening and relaxation of the chain-tuner, while the eigenfrequency of the cell was monitored in parallel, was chosen. This way, it was possible to monitor elastic and plastic deformation of the RF cell. 

\begin{figure}[!htbp]
	\centering
		\includegraphics[width=0.9\columnwidth]{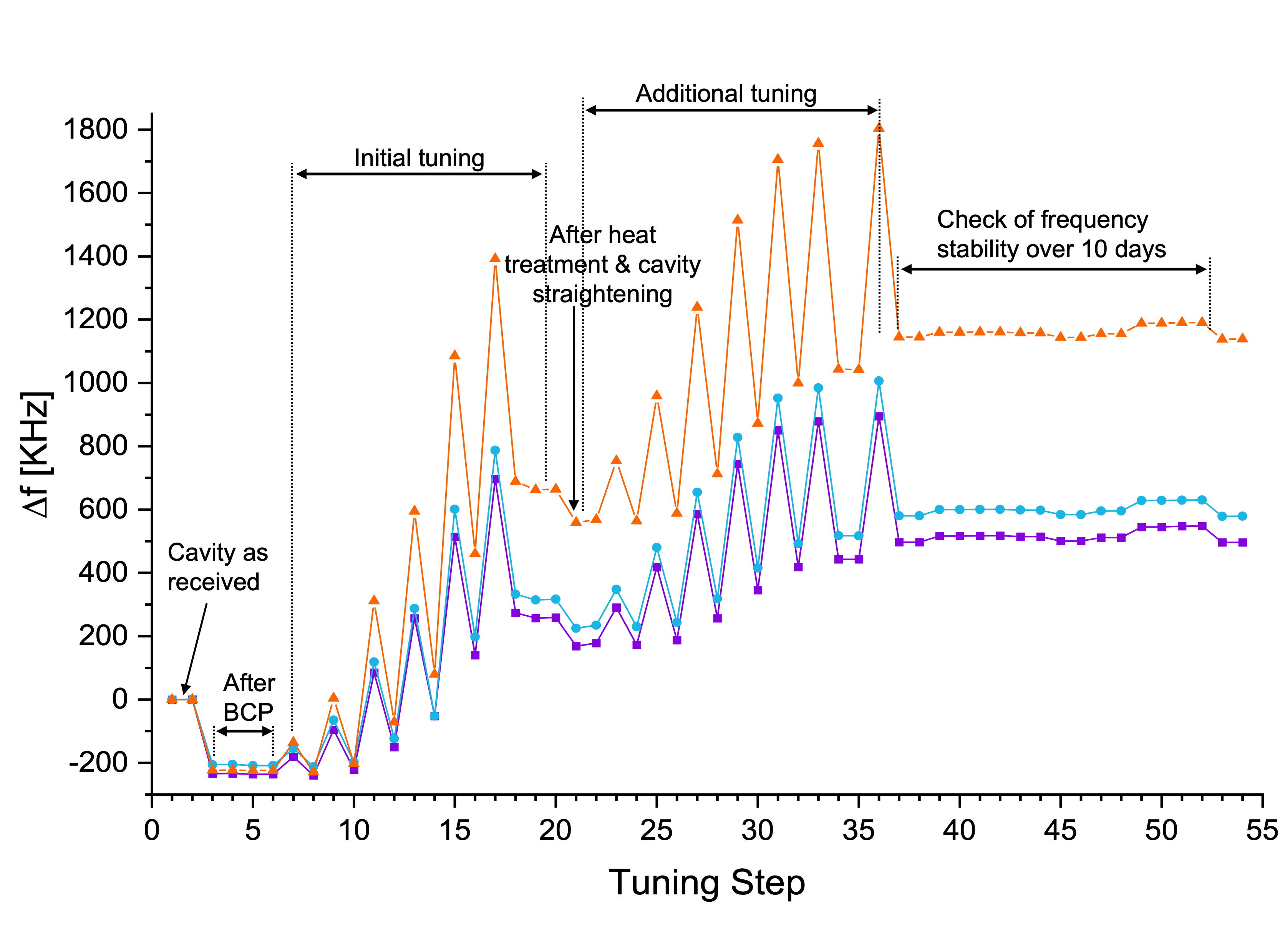}
	\caption{Frequency change vs. tuning steps for the three eigenmodes in the 2060-2100 MHz band. The orange line depicts the relevant TE\textsubscript{011} in cell 1, the other colours the two lower eigenmodes. The alternating tightening and relaxing approach used during the tuning procedure is clearly visible. A final increase of +1.2\,MHz for the relevant mode is achieved.}    	
	\label{fig:Tuning}
\end{figure}

\begin{figure}[!htbp]
	\centering
	
            \includegraphics[width=0.8\columnwidth]{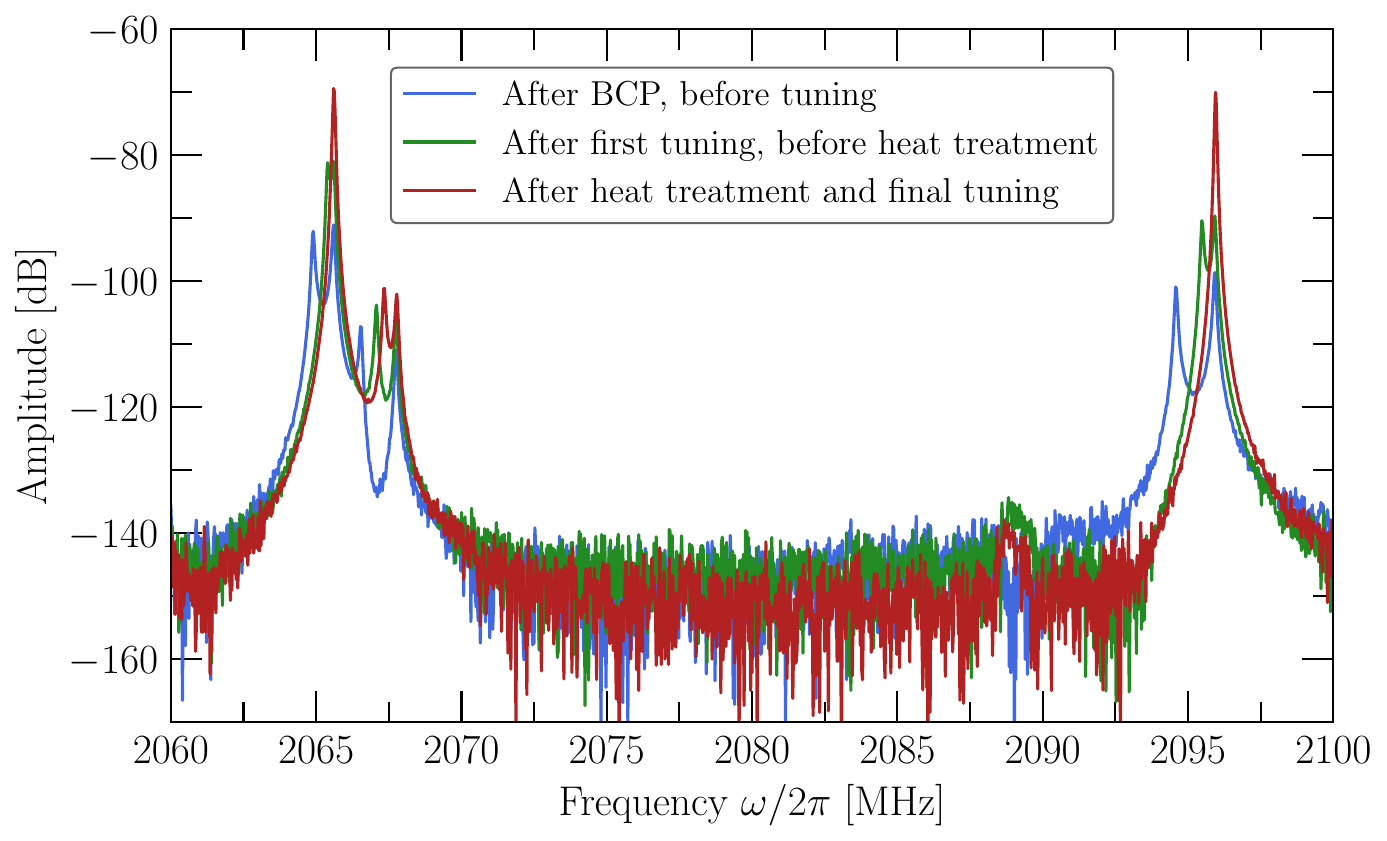}
	\caption{RF spectrum of the cavity at room temperature. The blue line shows the spectrum as measured on the cavity after the flash BCP, before any tuning step; the double peak structure of the TE\textsubscript{011} mode at 2096\,MHz is clearly visible. The green line shows the spectrum after $+$\SI{0.9}{\mega\hertz} of tuning and before the heat treatment (at step 21-26 from figure~\ref{fig:Tuning}), the TE\textsubscript{011} mode at 2096\,MHz is still split. The red line shows the spectrum after the tuning is completed. The two peaks merge and are not distinguishable anymore at room temperature.}    
	\label{fig:FinalSpectra}
\end{figure}

Figure~\ref{fig:Tuning} shows the cell 1 frequency progression starting from the arrival of the cavity at Fermilab up to the completion of the plastic tuning of the two cells. A frequency shift by 200\,kHz was introduced by the inner surface removal of $5\,\mu\text{m}$, which results in a sensitivity of 40\,kHz/$\mu$m, which is in the same order of magnitude of the sensitivity value (20\,kHz/$\mu$m) obtained from the simulations. Cell 1 was then tuned in multiple steps and an increase of the eigenfrequency by +1.2\,MHz was achieved. As this was more than intended, cell 2 was then also tuned by +200\,kHz. The following measurements (steps $>$ 44) were carried out to assess the stability of the cell tuning over 10 days. Small frequency variations are due to day to day temperature changes in the RF laboratory.
The final result in figure~\ref{fig:FinalSpectra} shows a comparison of the RF spectra before and after the room temperature tuning. 
After tuning, the two peaks merge due to the low quality factor at room temperature, which causes the resonances to be broad and overlap.
Measuring the frequency of each cell individually, as uncoupled system, shows that the frequency difference at room temperature is now on the order of (4-7)\,kHz. 

\section{Coupling coefficients}
\label{sec:couplings}

Gravitational waves have two distinct ways of coupling energy into the heterodyne cavity system, either by deforming and shaking its mechanical structure or by directly driving a transition from the loaded pump mode to an initially empty signal mode. The latter contribution has been shown to be small at frequencies below $\omega_g \sim \SI{1}{GHz}$ \cite{Berlin:2023grv} and is thus neglected in this discussion. Therefore, all calculations have been performed within the long-wavelength limit, where $L_\text{cav}\simeq V_\text{cav}^{1/3} \ll c/\omega_g$ for the effective length of the MAGO cavity $L_\text{cav}\sim \SI{0.5}{m}$. The mechanical interaction is characterised by two dimensionless interaction coefficients describing the excitation efficiency of the $m^\text{th}$ mechanical eigenmode by a GW $\Gamma_m$ and the energy transfer efficiency of mechanical mode $m$ between pump and signal mode $C_{01}^m$. 

\subsection{GW-mechanical mode coefficient}
\label{sec:GWmechCpl}

\begin{figure}[!htbp]
	\centering

 \subfigure[]{\includegraphics[height=5.1cm, width=0.49\textwidth]{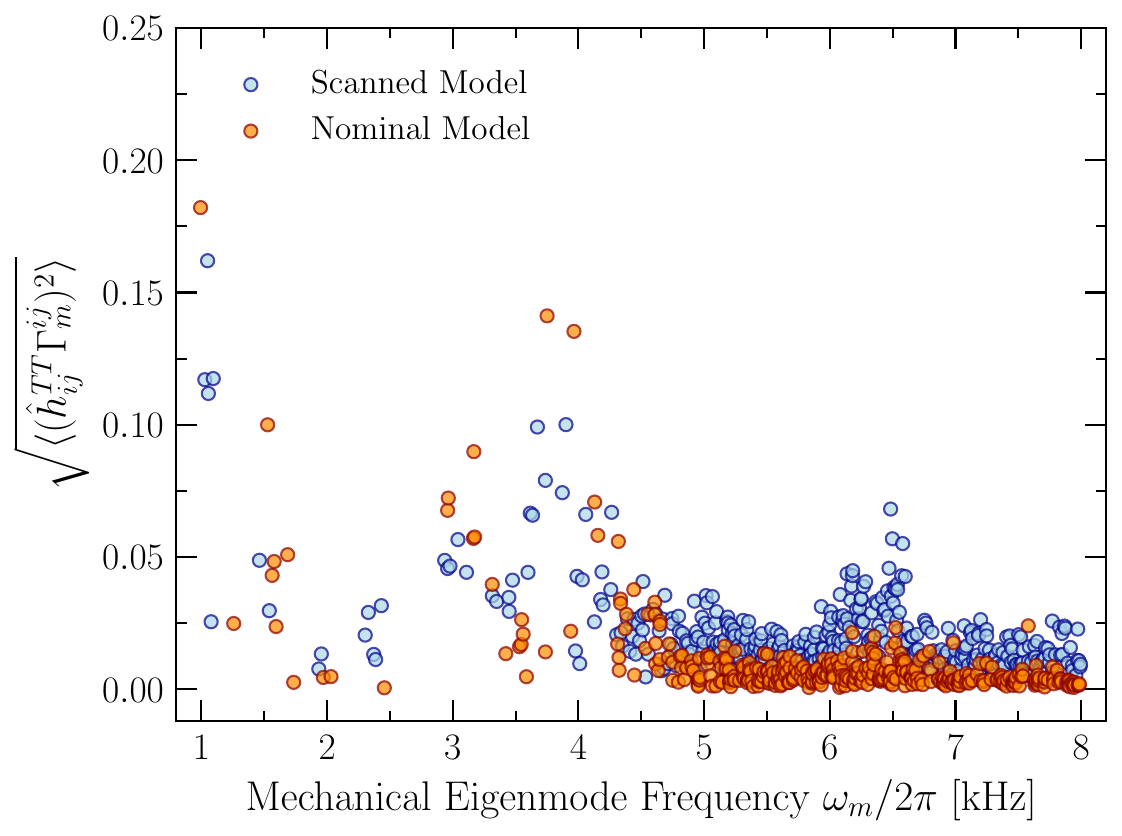}} 
    \subfigure[]{\includegraphics[height=5cm, width=0.49\textwidth]{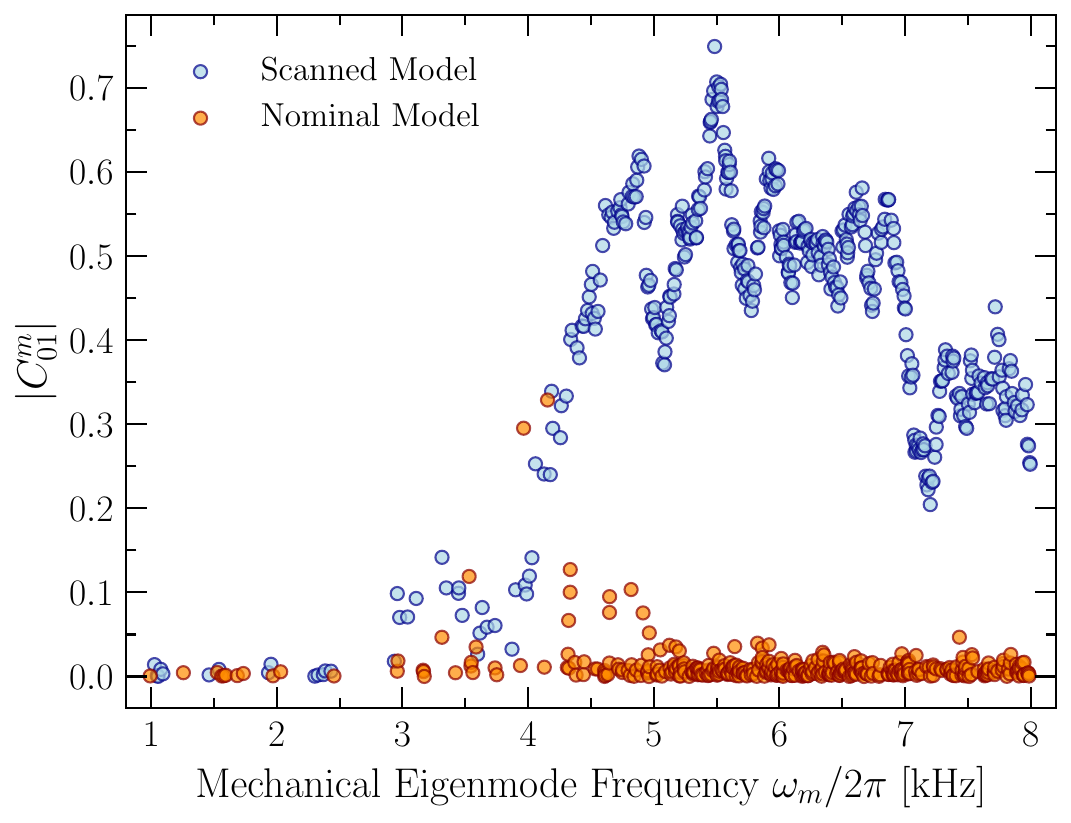}} 
        \subfigure[]{\includegraphics[height=5cm, width=0.49\textwidth]{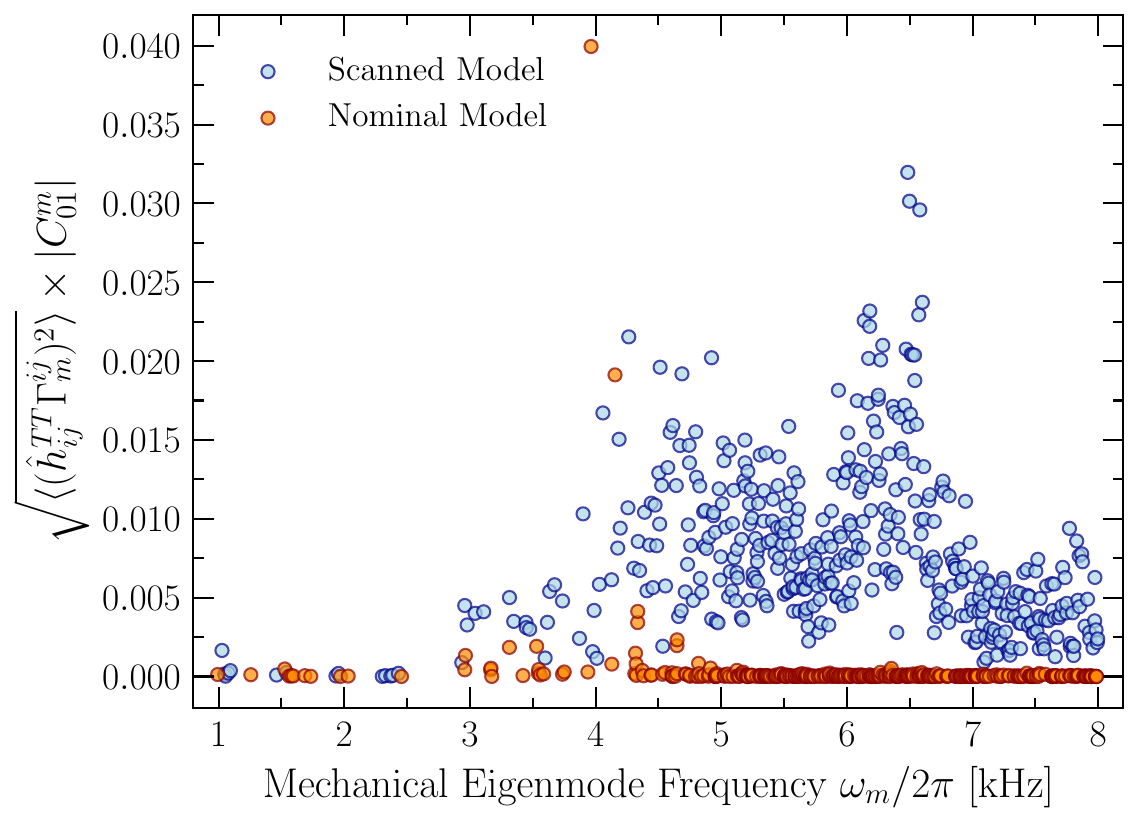}} 

	\caption{(a) Numerically evaluated coefficients $\Gamma_{m}$ averaged over direction and polarisation of an incoming GW coupling to the mechanical mode $m$ with eigenfrequencies $\omega_m/2\pi$. (b) Numerical evaluation of the internal coupling $C_{01}^m$ between the mechanical eigenmode $m$ and the electromagnetic modes. (c) Product of the two coupling coefficients. The nominal model exhibits a single mechanical mode that strongly contributes to the signal ($\omega_m /2\pi \simeq \SI{4}{kHz}$), whereas the scanned geometry shows multiple modes that are contributing but mostly weakly coupled. 
 }    	
	\label{fig:GWmechCplNumeric}
\end{figure}

For the parametrisation of an incoming periodic GW signal we use the strain tensor $h_{ij}^\text{TT} = h_0 R^{ik}R^{jl}\hat{h}_{kl}^\text{TT} e^{i\omega_g t}$ with $h_0$ the characteristic strain amplitude, $\omega_g$ the GW frequency, $R_{ij}=R_{ij}(\hat{\bf{n}}, \phi_p)$ a rotation matrix determining the direction $\hat{\bf{n}}$ of the GW and the polarisation $\phi_p$, and $\hat{h}_{ij}^\text{TT}=\text{diag}(1,-1,0)$. The force density acting on the mechanical structure in the proper detector (PD) frame is then given by $f_i = -1/2\rho(\Vec{x}) \Ddot{h}_{ij}^\text{TT} x^j$ \cite{maggiore:2007GWvol1}, where $\rho$ is the mass density function of the cavity walls. By decomposing the displacement of the cavity wall caused by $f_i$ into the mechanical eigenmodes of the cavity as $\vec{u}(\vec{x}, t) = \sum_m \vec{\xi}_m(\vec{x}) q_m(t)$ with $\int_{V_\text{cav}}dV\rho(\vec{x})|\vec{\xi}_m|^2=M$ the equation of motion for a mode $m$ is 
\begin{equation}
    \Ddot{q}_m + \frac{\omega_m}{Q_m} \Dot{q}_m + \omega_m^2 q_m = \frac{f_m}{M},
\end{equation}
where $M$ is the cavity mass, $\omega_m$ the mechanical eigenfrequency and $Q_m$ the mechanical quality factor of mode $m$. The generalised force density is hence given by 
\begin{equation}
    f_m(t) = -\frac{1}{2}\omega_g^2 V_\text{cav}^{1/3}M \hat{h}_{ij}^\text{TT} \Gamma^{ij}_m e^{i\omega_gt}
\end{equation}
with the cavity volume $V_\text{cav}$ and the dimensionless and scale independent GW-mechanical coupling 
\begin{align}
    \Gamma^{ij}_m = \frac{1}{V_\text{cav}^{1/3}M} \int_{V_\text{cav}} d^3x\, \rho(\vec{x}) x^i\cdot\xi_m^j(\vec{x}). 
\end{align}
We have numerically evaluated the $\Gamma_m^{ij}$ coupling coefficients with simulations in COMSOL\texttrademark\ covering the mechanical eigenmodes in the frequency range $1-\SI{8}{kHz}$. In figure \ref{fig:GWmechCplNumeric}(a) the direction and polarisation averaged values of $\Gamma^{ij}_m$ are shown for the eigenfrequencies $\omega_m/2\pi$. 
General considerations \cite{lobo:1995what, Berlin:2023grv} suggest that GWs most efficiently transfer energy to the quadrupole modes of a sphere. However, for overall geometries like the MAGO detector the coupling coefficient $\Gamma^{ij}_m$ is a measure of how much quadrupole moment the mechanical mode $m$ has. 
We expect that a few mechanical modes will dominate the coupling, primarily allowing the GW signal to couple to the cavity. From the numerical analysis we find that for both the ideal and the scanned MAGO model the maximal GW-mechanical coupling is of order $\mathcal{O}(0.1)$. 

\subsection{Mechanical-EM mode coefficient}

As shown in more detail in \cite{Berlin:2023grv, löwenberg} the change of boundary conditions due to the gravitational deformation of the cavity shell generates a mixing between the loaded pump mode and the empty signal mode, transferring energy between the two. After a decomposition of the electromagnetic fields inside the cavity $\Vec{E}(t,\Vec{x}) = \sum_n e_n(t) \Vec{E}_n(\Vec{x})$, where $\Vec{E}_n$ is the spatial pattern and $e_n(t)$ the time-dependent amplitude of the pump ($n = 0$) and signal ($n = 1$) mode, the equation of motion for the signal mode is 
\begin{equation}\label{eq:eomEfield}
    \Ddot{e}_1 + \frac{\omega_1}{Q_1} \Dot{e}_1 + \omega^2_1e_1 = V_\text{cav}^{-1/3} \sum_m q_m \sqrt{\frac{U_0}{U_1}}\omega_1^2 C_{01}^m e_0. 
\end{equation}
In equation \ref{eq:eomEfield} we assumed that $\omega_0\simeq\omega_1$, $U_n$ is the average energy in mode $n$ given by $U_n = \epsilon_0/2\int_{V_\text{cav}} d^3x|\Vec{E}_n|^2$, and we have defined the mechanical-EM overlap factor $C_{01}^m$ as in \cite{löwenberg, Ballantini:2003nt}
\begin{equation}\label{eq:overlapFactorDispEM}
    C_{01}^m = \frac{V_\text{cav}^{1/3}}{2\sqrt{U_1U_0}} \int_{\partial V_\text{cav}} d\vec{S} \cdot\vec{\xi}_m(\vec{x}) \bigg(\frac{1}{\mu_0} \vec{B}_1^*\cdot\vec{B}_0 - \epsilon_0\vec{E}_1^*\cdot\vec{E}_0\bigg).
\end{equation}
A numerical evaluation of the coupling coefficients $C_{01}^m$ for mechanical modes $m$ in the frequency range $1-\SI{8}{kHz}$ between the symmetric and anti-symmetric $\text{TE}_{011}$ electromagnetic mode of the MAGO cavity yield the results shown in figure \ref{fig:GWmechCplNumeric}(b). Note that due to the fixed polarisation of the modes in the MAGO geometry the $C_{01}^m$ coefficient does not depend on the direction of the GW.

Modes that contribute to the noise but are not excited by a GW are particularly crucial for the experimental setup \cite{Ballantini:2005am}. Such mechanical eigenmodes have $\Gamma_m\ll1$ and $C_{01}^m \neq 0$ while their frequency $\omega_m$ is close to the signal frequency $\omega_g$. If we compare the coupling coefficient product of the nominal and the scanned model in figure \ref{fig:GWmechCplNumeric}(c), the suppression of higher modes that is present in the nominal geometry is lost in the scanned detector model. In particular figure \ref{fig:GWmechCplNumeric}(b) and (c) show that all mechanical modes of the scanned model within the simulated range above $\SI{4}{kHz}$ contribute to the signal mode with $C_{01}^m$ of order $\mathcal{O}(1)$. 

\subsection{Noise budget estimate for MAGO cavity}
\label{sec:psd}
The goal of investigating the properties of the MAGO cavity in detail is to predict the GW sensitivity of our detector and identify the key parameters for an optimal setup. In order to characterise the sensitivity performance, we use the noise strain power spectral density (PSD)
\begin{equation}
    S_n(\omega_g)\coloneqq\frac{S_\text{noise}(\omega_0+\omega_g)}{|T(\omega_g)|^2},
\end{equation}
where $T$ is the \emph{transfer function} of our detector which describes the signal response $S_\text{sig}$ to a GW input $S_h$ with $S_\text{sig}(\omega_0+\omega_g)=|T(\omega_g)|^2S_h(\omega_g)$. Therefore, $S_n$ can be interpreted as the PSD of the noise had it entered the detector as a false GW strain i.e. $\text{sig}=h+n$. This means that the minimal GW strain that we naively expect to be resolvable in our output should be $h_\text{min}(\omega_g)\sim \sqrt{S_n(\omega_g)}$. However, it should be emphasised that the equivalent noise strain is agnostic of the integration time and data analysis techniques which can significantly increase the sensitivity depending on what kind of signal is considered. 

Far from a mechanical resonance of the cavity, the absolute value of the transfer function is given by
\begin{equation}\label{eq:transfer_function}
|T(\omega_g)|^2=\frac{\beta_\text{in}\beta_\text{out}}{(1+\beta_\text{in})^2}\frac{\omega_0}{Q_\text{int}}\,V_\text{cav}B_\text{eff}^2\,\big|C_{01}^m\Gamma_m\big|^2\,\frac{\omega_1^4}{(\omega_1^2-(\omega_0+\omega_g)^2)^2+\left(\frac{(\omega_0+\omega_g)\omega_1}{Q_1}\right)^2}\,,
\end{equation}
where we use $B_\text{eff}^2\coloneqq \frac{1}{V_\text{cav}}\int_{V_\text{cav}}\vec{B}^2_0$ as a measure of how efficiently the pump mode stores its energy\footnote{The quench limit of superconducting niobium is given by the superheating field $B_{sh}$ and requires $B_\text{eff}$ to be no more than $\mathcal{O}(0.2 \,\text{T})$.} and a sum of the coefficients $C_{01}^m\Gamma_m$ over all vibrational modes $m$ is implicit.$Q_\text{int}$ is the unloaded quality factor of the signal mode, only representing losses in the cavity walls. The parameters $\beta_\text{in}$ and $\beta_\text{out}$ describe how much power enters and exits the system compared to the power lost in the walls.
Details on the derivation of the transfer function and the following noise spectra can be found in \cite{Ballantini:2005am, Berlin:2023grv, löwenberg}. The transfer function near a mechanical resonance is more subtle, however the gravitational wave sensitivity is not necessarily enhanced. This is because vibrational noise sources are equally on resonance and the back-action of the electromagnetic fields on the cavity walls is further damping the oscillation \cite{löwenberg}.

\begin{figure}[!htbp]
	\centering
            \includegraphics[width=0.8\columnwidth]{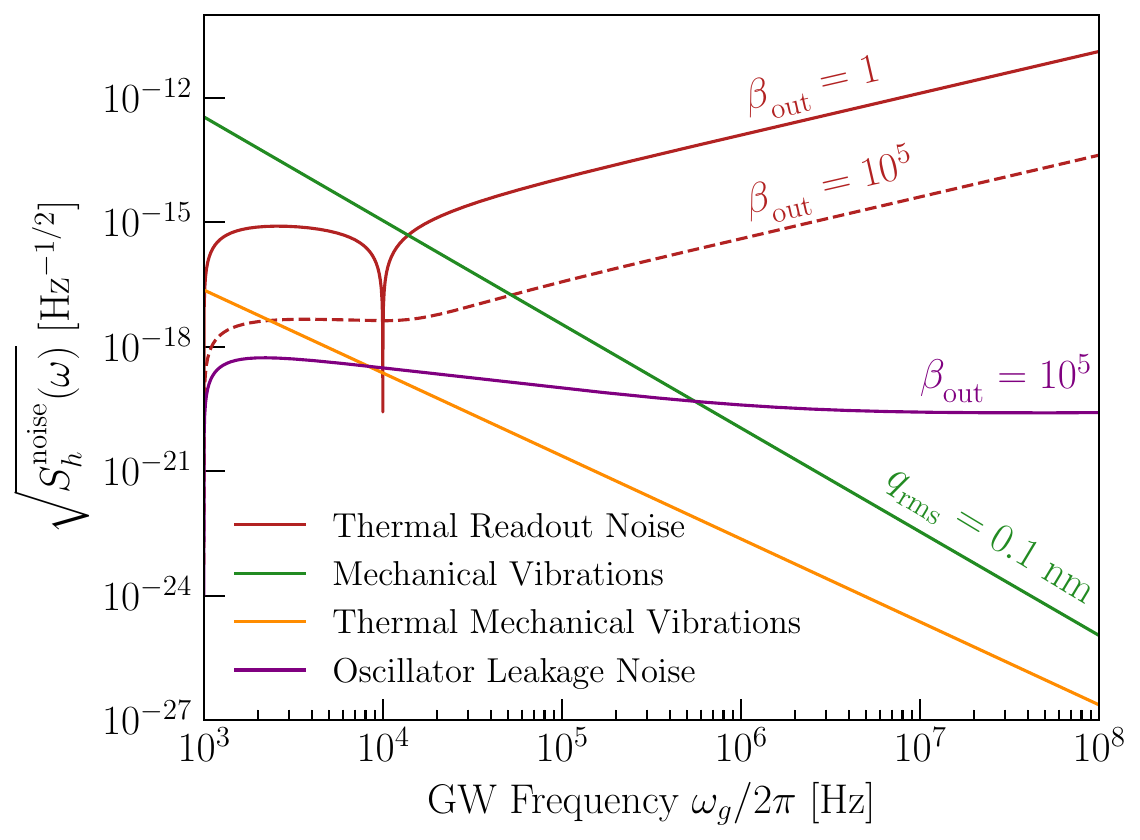}
	\caption{The three most relevant equivalent noise strain PSDs for the MAGO prototype cavity. The mechanical vibrations are estimated using equation \eqref{eq:mech_vib_noise} with $\omega_\text{ref}=10^3$ Hz and $\alpha = 1$. For the thermal vibrations equation \eqref{eq:th_vib_noise} is used. The thermal readout noise is shown using equation \eqref{eq:th_noise} where the $\omega_g$ dependence near resonance is using the unapproximated expression from equation \eqref{eq:transfer_function}. Furthermore, $\beta_\text{in}=1$ and $\Gamma_m=C_{01}^m=0.3$.}    	
	\label{fig:noise_estimate}
\end{figure}

The most relevant noise for a first test measurement with the MAGO cavity is expected to come from mechanical vibrations near the resonant frequency difference $\omega_g\approx\omega_1-\omega_0$ and from thermal noise in the readout system for $\omega_g\gg\omega_1-\omega_0$.
Thermal readout noise is not filtered by the cavity resonance and has a constant power spectrum $\propto k_B T$. For the equivalent noise strain we find 
\begin{equation}\label{eq:th_noise}
    \sqrt{S_{h_\text{th}}(\omega_g)}\approx 4.2\cdot 10^{-16}\,\text{Hz}^{-\frac12}\,\frac{1+\beta_\text{in}}{\sqrt{\beta_\text{in}\beta_\text{out}}}\left(\frac{Q_\text{int}}{10^{10}}\right)^{\frac12}\frac{B_\text{eff}}{0.1\,\text{T}}\,\frac{1}{C_{01}^m\Gamma_m}\,\frac{\omega_g-(\omega_1-\omega_0)}{10^5\,\text{Hz}}\,.
\end{equation}
Figure \ref{fig:noise_estimate} shows readout noise for critical coupling $\beta_\text{out}=1$ and overcoupling $\beta_\text{out}\gg1$. By reading out a larger fraction of the power in the cavity i.e. overcoupling we can increase the sensitivity where readout noise dominates and thus increase the bandwidth. However, it has to be noted that such strong couplings can affect the frequencies and fields in the cavity and thus change the coupling coefficient $C_{01}^m$ and maximal pump field $B_\text{eff}$.

Assuming that the vibrational noise is dominated by a single dominant quadrupole mode ($m=\text{quad}$), we find the equivalent strain for mechanical noise
\begin{equation}\label{eq:mech_vib_noise}
     \sqrt{S_{h_\text{mech.\ vib.}}(\omega_g)}\approx3.3\cdot10^{-14}\,\text{Hz}^{-\frac12}\frac{1}{\Gamma_\text{quad}}\frac{q_\text{rms}}{0.1\,\text{nm}}\left(\frac{10^6}{Q_\text{ref}}\right)^{\frac12}\left(\frac{\omega_\text{ref}}{\omega_g}\right)^{\frac{3+\alpha}{2}}\left(\frac{10^4\,\text{Hz}}{\omega_g}\right)^{\frac12}\,,
\end{equation}
where $\omega_\text{ref}$ is the frequency of a reference mode at which the RMS displacement of the cavity walls $q_\text{rms}$ can be measured or estimated. Away from this frequency, the noise source spectrum is extrapolated according to some power law $S_\text{mech. sources}\sim\omega_g^{-\alpha}$ where e.g.\ $\alpha = 1,\,2$ are reasonable choices \cite{Berlin:2023grv}. 

The exact value of $q_\text{rms}$ depends on the performance of the isolation and suspension system of the cavity. However, thermally excited vibrations pose an irreducible noise background which has successfully been reached by Weber Bar detectors in the past \cite{AURIGA}. For the MAGO prototype cavity it is found to be
\begin{equation}\label{eq:th_vib_noise}
     \sqrt{S_{h_\text{th.\ vib.}}(\omega_g)}\approx7.0\cdot10^{-20}\,\text{Hz}^{-\frac12}\frac{1}{\Gamma_\text{quad}}\left(\frac{10^6}{Q_\text{quad}}\right)^{\frac12}\left(\frac{10^4\,\text{Hz}}{\omega_g}\right)^{2}\,,
\end{equation}
where $Q_\text{quad}$ is the mechanical quality factor of the dominant quadrupole vibration. 

A last noise source to consider is given by phase and amplitude fluctuations in the oscillator driving the pump mode which can create sidebands at the signal frequency $\omega_1$. However, by utilising the symmetry difference between pump and signal mode in the cavity drive and readout system, this noise source can be geometrically suppressed by at least a factor $\epsilon\sim10^{-7}$ as described in \cite{Bernard:2000pz}. Furthermore, near mechanical resonances $\omega_1-\omega_0\simeq\omega_m$ this can lead to back-action effects on the mechanical oscillation but is negligible in the case of non-resonant wall vibrations which we are considering. Assuming the total noise PSD $S_\text{osc}$ including phase and amplitude noise of a commercially available oscillator \cite{Oscillator}, we find
\begin{equation}
    \sqrt{S_{h_\text{leak}}(\omega_g)}\approx4.5\cdot10^{-20}\,\text{Hz}^{-\frac12}\,\frac{\epsilon}{10^{-7}}\frac{\beta_\text{out}}{10^5}\frac{10^{10}}{Q_\text{int}}\frac{1}{C_{01}^m\Gamma_m}\frac{\omega_0+\omega_g}{\omega_1}\frac{\sqrt{S_\text{osc}(10^4\,\text{Hz})}}{1.2\cdot10^{-7}\,\text{Hz}^{-\frac12}}\,.
\end{equation}

These parametric estimates for the noise budget of the MAGO prototype are summarised in figure \ref{fig:noise_estimate}. 
A more detailed sensitivity estimate is possible as soon as the exact vibrational spectrum of the cavity inside its final support structure has been measured. Then, the exact background of mechanical excitations from the environment can be determined as well. Furthermore, the couplings $\beta_\text{in}$, $\beta_\text{out}$ and $B_\text{eff}$ will be known as soon as the dedicated RF system has been finalised.

\section{Conclusions}
\label{sec:conclusion}
This work constitutes a first step in the revival of the experimental R\&D programme for heterodyne detection of high-frequency GWs with microwave cavities. 
Further studies are underway to experimentally determine the optimal coupling to the pump and signal modes  and assess the impact of phase and amplitude noise in the presence of back-action effects as a function of mode overcoupling, and develop strategies for their mitigation. Furthermore, we will experimentally characterize the mechanical resonances of the detector and compare them with our simulations to improve on the rough estimate presented in the last chapter. Many key elements and challenges will also be directly relevant for the same search concept for axions~\cite{berlin_PhysRevD.104.L111701, berlin2020axion, giaccone2022design}.  
The {\tt PACO-2GHz-variable} cavity exhibits considerable deviations from the nominal geometry. This and fabrication accuracies necessitated a plastic tuning of the cavity, which likely has to be continued in some form also under cryogenic conditions, where the mode resolution is expected to improve strongly. Nevertheless, current measurements and simulations indicate, that the cavity can be used for a broadband search of GWs in the kHz to MHz regime, with the prospect to achieve sensitivities as expected. 

The next step in the commissioning of the cavity involves conducting RF tests at 2\,K. This phase is essential for evaluating the cavity's performance at cryogenic temperatures. During this process, we will study the field amplitudes within the cavity cells and analyse the RF spectrum at low temperatures, particularly as the cavity undergoes changes during the cooldown phase.
Additionally, we plan to measure the mechanical resonances of the cavity and the mechanical quality factor of niobium at cryogenic temperatures—a study that has not yet been undertaken with this cavity. Understanding these properties is crucial for optimising cavity performance and enhancing sensitivity.
Furthermore, we will investigate the RF control mechanisms and conduct a sensitivity study within the existing cryostat and suspension system. This preparatory work is intended to pave the way for a successful first physics run with the {\tt PACO-2GHz-variable} cavity in the existing cryostat, scheduled over the next two years. This measurement would result in the first GW constraint within the kHz to MHz range, though it is expected to be relatively weak. The long-term goals include developing an improved cavity design, engineering a dedicated low-noise cryostat and suspension system to significantly improve the sensitivity, and ultimately establish coordinated high-frequency GW observatories at DESY, Fermilab and possibly further locations.

\acknowledgments
For the loan of the {\tt PACO-2GHz-variable} cavity, we thank the Istituto Nazionale di Fisica Nuclare, Italy.
The authors thank Asher Berlin, Julien Branlard, Sergio Calatroni, Andrea Chincarini, Sebastian Ellis, Gianluca Gemme, Roni Harnik, Frank Ludwig, Linus Pfeiffer, Andreas Ringwald, Holger Schlarb, Udai Raj Singh, Louise Springer and Hans Weise for their support and useful discussions. 
This material is based upon work supported by the U.S. Department of Energy, Office of Science, National Quantum Information Science Research Centers, Superconducting Quantum Materials and Systems Center (SQMS) under contract number DE-AC02-07CH11359. LF, WH, TK, GM-P, KP and MW acknowledge support by the BMBF under the research grant 05H21GURB2 and the project is also funded/acknowledges support by the Deutsche Forschungsgemeinschaft (DFG, German Research Foundation) under Germanys Excellence Strategy - EXC 2121 `Quantum Universe' - 390833306.



\bibliographystyle{JHEP}
\bibliography{biblio.bib}

\end{document}